\pdfoutput=1
\documentclass[pre,twocolumn,showpacs,superscriptaddress,aps]{revtex4}

\usepackage{amsfonts}
\usepackage{amsmath}
\usepackage{amssymb}
\usepackage{graphicx}
\usepackage{dcolumn}
\usepackage{times}
\usepackage{color}

\DeclareMathOperator{\sech}{sech}

\begin{document}

\title{Coupled backward- and forward-propagating solitons in a composite
right/left-handed transmission line}
\author{G. P. Veldes}
\affiliation{Department of Physics, University of Athens, Panepistimiopolis,
Zografos, Athens 15784, Greece}
\affiliation{Department of Electronics, Technological Educational Institute of Lamia, Lamia 35100,Greece}
\author{J. Cuevas}
\affiliation{Grupo de F\'{\i}sica No Lineal, Universidad de Sevilla. Departamento de F\'{\i}sica Aplicada I,
Escuela Polit\'ecnica Superior, C/ Virgen de \'{A}frica, 7, 41011
Sevilla, Spain}
\author{P. G.\ Kevrekidis}
\affiliation{Department of Mathematics and Statistics, University of Massachusetts,
Amherst MA 01003-4515, USA}
\author{D. J.\ Frantzeskakis}
\affiliation{Department of Physics, University of Athens, Panepistimiopolis,
Zografos, Athens 15784, Greece}

\begin{abstract}

We study the coupling between backward- and forward-propagating wave modes,
with the same group velocity, in a composite
right/left-handed nonlinear transmission line. Using an asymptotic multiscale expansion
technique, we derive a system of two coupled nonlinear Schr{\"o}dinger equations governing the
evolution of the envelopes of these modes. We show that this system supports a variety
of backward- and forward propagating vector solitons, of the bright-bright, bright-dark and
dark-bright type. Performing systematic numerical simulations in the framework of the
original lattice that models the transmission line, we study the propagation properties of
the derived vector soliton solutions. We show that all types of the predicted solitons exist,
but differ on their robustness: only bright-bright solitons propagate undistorted for long times,
while the other types are less robust, featuring shorter lifetimes. In all cases, our analytical
predictions are in a very good agreement with the results of the simulations, at least up to
times of the order of the solitons' lifetimes.

\end{abstract}

\pacs{41.20.Jb, 42.65.Tg, 78.20.Ci}

\maketitle

\section{Introduction }

Left-handed (LH) metamaterials are artificial, effectively homogeneous structures, featuring
negative refractive index at specific frequency bands where the
effective permittivity $\epsilon$ and permeability $\mu$ are simultaneously negative
\cite{review1,review2,review3}. In fact, all known realizations of LH metamaterials rely on
the use of common right-handed (RH) elements and, thus, in a realistic situation such a
composite material features both a LH and a RH behavior, in certain frequency bands.
Physically speaking, the difference between the two is that in the LH (RH) regime,
the energy and the wave fronts of the electromagnetic (EM) waves
propagate in the opposite (same) directions, giving rise to backward- (forward-)
propagating waves.

Transmission line (TL) theory constitutes a convenient framework for
the analysis of LH metamaterials. Such an
analysis relies on the connection of the EM properties of the medium ($\epsilon$ and
$\mu$) with the electric elements of the TL's unit cell, namely the serial and shunt
impedance. As mentioned above, in practice {\it composite right/left-handed (CRLH)}
structures are quite
relevant, giving  rise to pertinent CRLH-TL models.
These models are, in fact,
dynamical lattices which can be used for the description of a variety of metamaterials-based
devices and systems, such as resonators, directional couplers, antennas, etc
\cite{review2,review1,review3,Caloz1}.

Nonlinear CRLH-TLs, with a serial or/and shunt impedance depending on voltages or currents, have
also attracted attention. Such structures may be realized by inserting diodes -- which mimic
voltage-controlled nonlinear capacitors -- into resonant conductive elements (such as split-ring
resonators) \cite{diodes,Carbonell,diodes2}. Such nonlinear CRLH-TL models have been used
in various works dealing, e.g., with the parametric shielding of EM fields \cite{shield},
the long-short wave interaction \cite{tatar}, or soliton formation \cite{solTL1,solTL2,solTL3}.
Experiments in nonlinear CRLH-TLs have also been performed (see the review \cite{revtl}),
and formation of bright \cite{BS,lars} or dark \cite{lars,DS} envelope solitons, described by an
effective nonlinear Schr\"{o}dinger (NLS) equation, was reported. Notice that in earlier
studies on RH-TL models it was shown that two (or more) solitons propagating with the same group
velocity, can be described by a system of two (or more) NLS equations \cite{inoue1} (see also
\cite{bilbault} for theoretical as well as experimental results). Such coupled NLS equations
have been studied extensively in nonlinear optics and mathematical physics;
see, e.g., Refs.~\cite{Remoissenet,kivagr,truba}
and references therein. They are well-known to
give rise to a variety of vector solitons,
including bright-bright, bright-dark, and dark-dark ones.

In this work, we study analytically and numerically the interaction between backward- and
forward-propagating solitons in a nonlinear CRLH-TL. Our model is a nonlinear version of
a generic CRLH-TL model (see, e.g., Refs.~\cite{review2,Caloz1}): the considered nonlinear element
in the unit cell of the TL is the shunt capacitor, which simulates the presence of a
heterostracture barrier varactor (HBV) diode \cite{Carbonell} (the capacitance of the HBV diode
depends on the applied voltage). Starting from the discrete lump element model of the CRLH-TL,
we derive a nonlinear lattice equation. First, we study the linear regime and show that,
for certain frequency bands, RH- and LH-modes can propagate with the same group velocity.
Next, we treat the nonlinear lattice equation in the framework of the {\it quasi-discrete}
(or {\it quasi-continuum}) approximation (see, e.g., \cite{Marquie,lars,veldes} and \cite{Remoissenet}
for a review): we thus seek for envelope soliton solutions of the nonlinear lattice model,
characterized by a {\it discrete carrier} and a {\it continuum envelope} and employ
an asymptotic multi-scale expansion method, to derive a system of two coupled NLS equations.
Each of these equations describes the evolution of the envelope of a backward- (LH-) and
a forward-propagating (RH-) mode. A systematic analysis of the system of the NLS equations
reveals the existence --in certain frequency bands-- of three different types of vector solitons:
(a) a backward-propagating bright soliton coupled with a forward-propagating bright soliton,
(b) a backward-propagating bright soliton coupled with a forward-propagating dark soliton.
(c) a backward-propagating dark soliton coupled with a forward-propagating bright soliton, and

The above analytical predictions are then tested against direct numerical simulations, which are
performed in the framework of the original nonlinear lattice model. The results of the simulations
verify the existence of the aforementioned types of vector solitons
in the full TL model, but also offer important
information regarding their robustness. In particular,
results of direct simulations performed for long times
indicate that bright-bright solitons
are the most robust among the members of the vector soliton family. Indeed,
the mixed (dark-bright or bright-dark) types are found to be less robust; however,
the dark-bright
solitons in a specific frequency band, although they are deformed during their evolution,
are found to be more robust than those in other bands, as well as the bright-dark solitons,
which are destroyed for the same propagation time.
In any case, our results
indicate the existence of all three types, robustness of bright-bright solitons and partial or substantial deformation
of the other types. We can thus conclude that
bright-bright (LH-RH), as well as dark-bright (LH-RH) solitons in certain frequency bands, have a better
chance to be observed in experiments.

The paper is organized as follows. In Section~II, we introduce the nonlinear CRLH-TL model and
the pertinent lattice equation, and derive the system of the two coupled NLS equations
(relevant details are also appended in an Appendix).
In Section~III, we present analytical and numerical results for each type of vector soliton.
Finally, in Section~IV, we summarize our conclusions.

\section{The model and its analytical considerations}

\subsection{The nonlinear CRLH-TL model}

\begin{figure}[tbp]
\centering
\includegraphics[width=8cm]{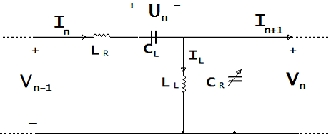}
\caption{The unit-cell circuit of the nonlinear CRLH model.}
\label{fig:circuitcrlh model}
\end{figure}

We consider a generic CRLH-TL, composed by both right- and left-handed elements, as shown in its
unit-cell circuit shown in Fig.~\ref{fig:circuitcrlh model} \cite{review2,Caloz1}.
The (RH) elements of this TL are the inductance $L_{R}$ and capacitance $C_{R}$, while the LH ones
are the inductance $L_{L}$ and capacitance $C_{L}$. We assume that the TL is loaded with a
nonlinear capacitance ($C_R$, while the capacitance $C_L$
will be assumed to be fixed and voltage independent).
 This can be implemented by proper insertion
of diodes in the TL (see, e.g., pertinent experiments as well as theoretical work in
Refs.~\cite{solTL1,solTL2,solTL3,revtl,BS,lars,DS}); in other words, we assume that the shunt
capacitor $C_R$ is nonlinear (see details below).

Let us now consider Kirchhoff's voltage and current laws for the unit-cell circuit of Fig.~\ref{fig:circuitcrlh model},
which respectively read:
\begin{eqnarray}
V_{n-1} &=& V_n+L_{R}\frac{dI_{n}}{dt}+U_n,
\label{KVL} \\
I_n &=& I_{n+1}+I_L +  \frac{d}{dt}(C_R V_n),
\end{eqnarray}
where $U_n$ is the voltage across the capacitance $C_L$ and
$I_L$ is the current across the inductor $L_L$. The above equations, together with the auxiliary equations
$V_n= L_L dI_L/dt$ and $I_n=C_L dU_n/dt$, lead to the following system:
\begin{eqnarray}
&& L_{R}L_{L}C_{L}\frac{d^{4}}{dt^{4}}(C_R V_n)+L_{L}\frac{d^{2}}{dt^{2}}(C_R V_n)+L_{R}C_{L}\frac{d^{2}V_n}{dt^{2}}\nonumber\\
&&-L_{L}C_{L}\frac{d^{2}}{dt^{2}}(V_{n+1}+V_{n-1}-2V_{n})+V_{n}=0.
\label{eq:eq01}
\end{eqnarray}

To proceed further, we now consider a specific voltage-dependence for the nonlinear capacitance $C_{R}$. Here, we will assume
that -- for sufficiently small values of the voltage $V_n$ -- the function $C_{R}(V_n)$ can be approximated as follows, via a
Taylor expansion:
\begin{eqnarray}
C_R(V_{n}) \approx C_{R0}+C_{R0}^{'}(V_n-V_0)+\frac{1}{2}C_{R0}^{''}(V_n-V_0)^{2},
\label{eq:Taylor1}
\end{eqnarray}
where $C_{R0} \equiv C_R(V_{0})$ is a constant
capacitance corresponding to the bias voltage $V_{0}$, while
$C_{R0}^{'}$
and $C_{R0}^{''}$
also assume constant values, depending on the particular form of $C_{R}(V)$.
Below, we will further discuss
this approximation, in connection with the HBV diode, used in the
experiments described in Ref.~\cite{revtl} (similar varactor-type diodes were also used in
the experiments of Ref.~\cite{diodes2}).

Next, substituting Eq.~(\ref{eq:Taylor1}) into Eq.~(\ref{eq:eq01}) and using the scale transformations
$t \rightarrow \omega_{sh} t$ [where $\omega_{sh}^{2}= (L_L C_{R0})^{-1}$] and
$V_n \rightarrow [C_{R0}^{'}(2C_{R0})^{-1}]V_n $, we obtain:
\begin{eqnarray}
&&\frac{d^{4}V_n}{dt^{4}}-\beta^{2}\frac{d^{2}}{dt^{2}}(V_{n+1}+V_{n-1}-2V_{n})+(1+\delta^{2})\frac{d^{2}V_n}{dt^{2}}\nonumber\\
&&+\delta^{2}V_{n}+\delta^{2}\frac{d^{2}V_{n}^{2}}{dt^{2}} +\delta^{2}\mu
\frac{d^{2}V_{n}^{3}}{dt^{2}} +\frac{d^{4}V_{n}^{2}}{dt^{4}}+\mu\frac{d^{4}V_{n}^{3}}{dt^{4}}=0,
\nonumber\\
\label{eq:model}
\end{eqnarray}
where the constant parameters $\delta$, $\beta$ and $\mu$ are given by:
\begin{eqnarray}
\delta=\frac{f_{\rm se}}{f_{\rm sh}}, \,\
\beta=\frac{f_{RH}}{f_{\rm sh}},\,\
\mu=\frac{2C_{R0}^{''}}{3C_{R0}^{'2}}C_{R0}.
\label{eq:param1}
\end{eqnarray}
In the above expressions, $f_{\rm se}$ and $f_{\rm sh}$ denote series and shunt frequencies, while $f_{RH}$ denotes the
characteristic frequency related to the RH part of the unit-cell circuit, respectively; the above frequencies are defined as:
\begin{eqnarray}
f_{\rm se}&=&\frac{1}{2 \pi \sqrt{L_{R}C_{L}}},\,\
f_{\rm sh}=\frac{1}{2 \pi \sqrt {L_LC_{R0}}},\,\,\nonumber\\
f_{\rm RH}&=&\frac{1}{2 \pi \sqrt {L_{R}C_{R0}}}.
\label{eq:freq}
\end{eqnarray}
Note that if $f_{\rm se}/f_{\rm sh}=1$, i.e., $\delta=1$, then the CRLH-TL is usually referred to as {\it balanced}, in the sense that the characteristic impedances of the purely LH- and RH-TL, defined as $Z_L=\sqrt{L_L/C_L}$ and $Z_R=\sqrt{L_R/C_{R0}}$, are equal, i.e., $Z_L=Z_R$ \cite{review2}. On the other hand, if $f_{\rm se}/f_{\rm sh}>1$, i.e., $\delta>1$, the LH part of the TL dominates, in the sense that the TL has a more pronounced LH behaviour (the serial branch features a capacitive character while the shunt branch an inductive one). In the opposite case, $f_{\rm se}/f_{\rm sh}<1$, i.e., $\delta<1$, the RH part of the TL dominates and the TL has a more pronounced RH behaviour (the serial branch features
an inductive character while the shunt branch a capacitive one).

It is now relevant to adopt physically relevant parameter values for Eq.~(\ref{eq:model}). For applications in the microwave
frequency range (e.g., for microstrip lines \cite{review2} or coplanar waveguide structures loaded with SRRs \cite{review3} --
cf. also Ref.~\cite{veldes} for recent work), typical values of the capacitances and inductances involved in the CRLH
structure are of the order of pF and nH, respectively. Here, we will use the values $L_{R}=1$~nH, $C_{L}=0.1$~pF, and
$L_{L}=0.12$~nH; thus, the frequencies in Eqs.~(\ref{eq:freq}) take the values
$f_{\rm se}=15.92$~GHz, $f_{\rm sh}=14.53$~GHz and
$f_{\rm RH}=5.03$~GHz. On the other hand, as concerns the parameters involved with the nonlinear capacitor $C_R$, we assume
that the pertinent capacitance corresponds to a HBV diode, which is characterized by the following equation \cite{Carbonell}
(see also \cite{diodes2}, where the same form of $C(V)$ is used, but different parameter values):
\begin{equation}
C(V)=C_{j0}A_{da} \left(1+\frac{|V|}{V_{br}} \right)^{-m},
\label{hbv}
\end{equation}
where $C_{j0} = 1.53$~fF/$\mu$m$^2$ is the capacitance corresponding to bias voltage $V_0=0.2$~V,
$A_{da} =650~\mu$m$^2$ is the device area, $V_{br}=12$~V is the breakdown potential, and the exponent
$m =2.7$ results from fitting experimental data.
It is clear that, for sufficiently small $V$, by Taylor expanding Eq.~(\ref{hbv}) one obtains
Eq.~(\ref{eq:Taylor1}), where the constant parameter values involved are
$C_{R0}=1$~pF,
$C_{R0}^{'} =-0.24$~pF/V and
$C_{R0}^{''} =-0.08$~pF/V$^{2}$. To this end, the values of the normalized parameters
$\delta$, $\beta$ and $\mu$ appearing in Eq.~(\ref{eq:model}) take the following values:
\begin{equation}
\delta \approx 1.1, \qquad \beta \approx 0.35, \qquad \mu \approx -0.9.
\label{param}
\end{equation}
Below, we will use these values for the purposes of our analytical and numerical considerations
(we have checked that other values lead to qualitatively similar results). Notice that our choice
leads to $\delta>1$, i.e., we consider the case where the TL has a more pronounced LH character;
however, when considering the linear setting (see next subsection), this parameter will
also assume other values, corresponding to the balanced and RH-dominated behaviour as well.

\subsection{Linear analysis}
\begin{figure}[tbp]
\begin{tabular}{cc}
\includegraphics[scale=0.4]{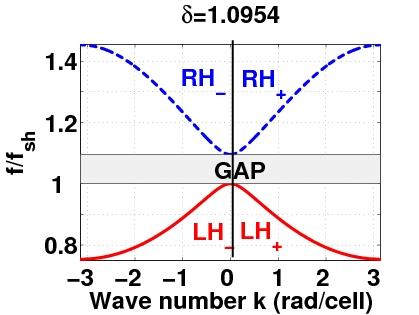}\\
\includegraphics[scale=0.4]{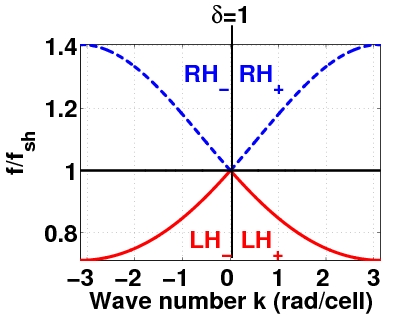}\\
\includegraphics[scale=0.4]{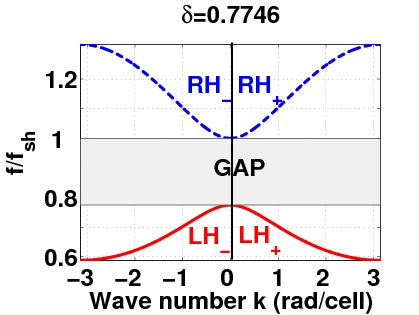}\
\end{tabular}
\caption{(Color online) The dispersion relation showing the normalized frequency $f/f_{\rm sh}$ as a function of the
wave number $k$ (in rad/cell) for different values of $\delta$, i.e., $\delta=1.0954$ (top panel),
$\delta=1$ (middle panel), and $\delta=0.7746$ (bottom panel). The solid (red) and dashed (blue) lines show the dispersion relation in the LH- and RH-frequency regions, respectively; RH$_{\pm}$ and LH$_{\pm}$ denote branches with $k>0$ or $k<0$. If $\delta \ne 1$ a gap is formed; the width of the gap is $|\delta-1|$ for $\delta>1$ (top panel) or $\delta<1$ (bottom panel).}
\label{fig:dispersion}
\end{figure}

We now assume plane wave solutions of Eq.~(\ref{eq:model}), of the form $V_n=V_{\rm o}\exp[i(kn-\omega t)]$, where $k$ and $\omega$ denote the wave number and angular frequency, respectively, while the amplitude of the wave is $V_{\rm o} \ll 1$. Substituting the above ansatz into Eq.~(\ref{eq:model}), and keeping
only the linear terms in $V_0$,
we obtain the following linear dispersion relation:
\begin{equation}
\omega^{4}-\left(1+\delta^2+4\beta^2\sin^{2}\frac{k}{2}\right)\omega^{2}+\delta^2=0.
\label{eq:linear_disp}
\end{equation}
The above result is illustrated in Fig.~\ref{fig:dispersion}, where we plot the frequency
$f/f_{\rm sh}$ as a function of the wave number $k$ (in rad/cell), for three different values of $\delta$
(note that here we consider one period of $k$, i.e., $-\pi \leq k_j \leq \pi$).
It is clear that for $\delta=1.0954$ (top panel) there exist two frequency bands where EM wave propagation
is possible: the RH-band [high-frequency band depicted by dashed (blue) line], for $1.0954<f<1.4535$, and the LH-band [low-frequency band depicted by solid (red) line], for $0.7538<f<1$.
In the same case ($\delta=1.0954$), there exists a gap for $1<f/f_{\rm sh}<\delta$,
where EM wave propagation is not possible.

In the case where $\delta=1$ (corresponding, e.g., to the value $C_{L}=0.12$~pF) the gap vanishes (cf. middle panel of Fig.~\ref{fig:dispersion}) and the TL is balanced. In the balanced case, EM wave propagation is possible in two frequency bands as well: the RH-band [high-frequency band -- cf. dashed (blue) line] with $1<f/f_{\rm sh}<1.405$ and the LH-band [low-frequency band -- cf. solid (red) line] with $0.7117<f/f_{\rm sh}<1$.

Finally, for $\delta= 0.7746$ (corresponding, e.g., to $C_{L}=0.2$~pF), a gap appears again for $\delta<f/f_{\rm sh}<1$ (bottom panel of Fig.~\ref{fig:dispersion}). In this case too, there exist a RH-frequency band and a LH-frequency band, for $0.588<f/f_{\rm sh}<0.7746$ and $1<f/f_{\rm sh}<1.317$, respectively. Note that in all cases, the RH$_{\pm}$ and LH$_{\pm}$ branches correspond to positive or negative $k$, respectively.

Thus, generally, in the linear setting -- and for a given frequency -- the EM waves may either propagate
in the RH region (forward wave propagation) or in the LH region (backward wave propagation). However,
in the nonlinear setting, coupling between modes propagating in the LH and RH regime is possible
(see, e.g., relevant earlier work in Refs.~\cite{inoue1,bilbault}). Below we will demonstrate that
this is the case indeed, and study the coupling (interaction) between LH and RH modes with equal
group velocities. Since the latter are tangents in the dispersion curves, inspection of
Fig.~\ref{fig:dispersion} shows that it is possible to identify domains, belonging to the
RH$_{\pm}$ and LH$_{\mp}$ branches, exhibiting parallel tangents, i.e., equal group velocities.

\begin{figure}[tbp]
\begin{tabular}{cc}
\includegraphics[scale=0.4]{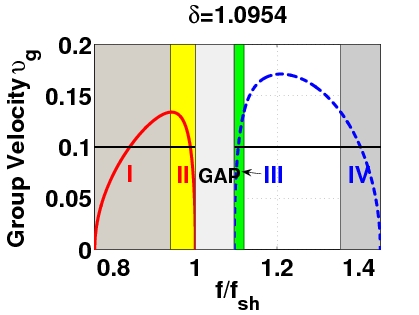}\\
\includegraphics[scale=0.4]{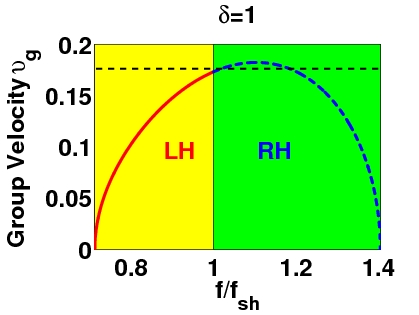}\\
\includegraphics[scale=0.4]{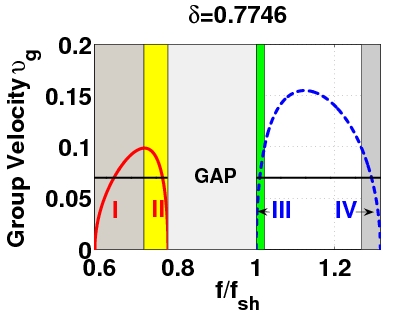}
\end{tabular}
\caption{(Color online) The group velocity $v_g$ as a function of the normalized
frequency $f/f_{\rm sh}$ (for $\delta=1.0954$). The solid (red) and dashed (blue) lines indicate
branches corresponding to the LH$_{-}$ and RH$_{+}$ regimes, respectively. The intersection of the
group velocity curves with the horizontal solid (black) line depicts frequencies of modes with the same
group velocity, $v_g=0.1$. Regions I, II, III and IV indicate
possible interactions between LH$_{-}$ and RH$_{+}$ modes with the same $v_g$ but different signs of
GVD.}
\label{fig:gvel}
\end{figure}

To further elaborate on this, we may use Eq.~(\ref{eq:linear_disp}) to obtain the group velocity
$v_{g} \equiv \partial \omega/\partial k$:
\begin{equation}
v_{g}=
\frac{\omega^3 \beta^2\sin k}{\omega^{4}-\delta^{2}}.
\label{eq:gvelqc}
\end{equation}
In Fig.~\ref{fig:gvel}, we show the dependence of the group velocity $v_g$ on the normalized frequency
$f/f_{\rm sh}$, for the values of $\delta$ used in Fig.~\ref{fig:dispersion}.
Notice that the figure depicts only the group-velocity branches
with $v_g>0$ -- see solid (red) and dashed (blue) lines -- corresponding, respectively, to the LH$_{-}$
and RH$_{+}$ branches of the dispersion curves;
the branches with $v_g<0$ (pertinent to the LH$_{+}$ and RH$_{-}$ branches of the dispersion curve) are
mirror symmetric with respect to the ones shown in the figure, due to the parity of the
dispersion relation.

Considering a horizontal cut of the group-velocity curves, say at $v_g=0.1$ or $v_g=0.075$ (see horizontal lines in
the top and bottom panels of Fig.~\ref{fig:gvel}), it is readily observed that, indeed,
a LH$_{-}$ and a RH$_{+}$ mode can share a common group velocity (and interact in the nonlinear regime,
as mentioned above). In fact, inspection of the group-velocity curves, say in the top panel of
Fig.~\ref{fig:gvel}, shows that the maximum possible common $v_g$ is given by $v_{g_{max}}=0.1339$,
the local maximum of $v_g$, occurring at $f=0.9391$, in the (shorter in height) LH
frequency band. Then, one can divide each of the LH and RH group-velocity curves into two sub-regions,
depending on the sign of the group-velocity dispersion (GVD), $\partial v_g/\partial \omega$, where
such coupling with equal group velocities may occur. These subregions are:
(a) the sub-bands I ($0.7538<f/f_{\rm sh}<0.9391$) and II ($0.9391<f/f_{\rm sh}<1$)
for the LH-frequency band, characterized by positive and negative GVD respectively, and
(b) the sub-bands III ($1.0954<f/f_{\rm sh}<1.1195$) and IV ($1.356<f/f_{\rm sh}<1.4535$) for the RH-frequency band,
again characterized by positive and negative GVD respectively. Thus, nonlinear LH and RH modes of equal $v_g$
can feature the following four different possible interactions:
\begin{enumerate}
\item {\it LH-mode in band II and RH-mode in band IV},
both featuring negative GVD.
\item {\it LH-mode in band I and RH-mode in band IV};
here, the LH (RH) mode features positive (negative) GVD.
\item {\it LH-mode in band I and RH-mode in band III},
both featuring positive GVD.
\item {\it LH-mode in band II and RH-mode in band III};
here, the LH (RH) mode features negative (positive) GVD.

\end{enumerate}

It is clear that the above set of possibilities arises from the existence of the gap in the considered case
with $\delta=1.0954$. A similar situation also occurs for $\delta <1$, e.g., for $\delta= 0.7746$ as in the
bottom panels of Figs.~\ref{fig:dispersion} and \ref{fig:gvel}.
On the other hand, for $\delta = 1$ the gap does not longer exist and,
thus, the only possible interaction is between a LH-mode with positive GVD and a RH-mode with negative GVD;
this interaction can occur for group velocities $v_g \le 0.175$, i.e., beneath the dashed horizontal line in
the middle panel of Fig.~\ref{fig:gvel}. This possibility, however, is already taken into
regard -- cf. case (2) above; furthermore, soliton formation in the balanced CRLH-TL ($\delta = 1$)
has already been studied in the literature \cite{solTL2}. For these reasons,
below we will proceed by analyzing the case corresponding to $\delta=1.0954$, which offers all possible scenarios;
it is clear that the case of $\delta=0.7746$ shares similar qualitative
features; this similarity extends beyond the linear wave case and into
the nonlinear solitonic one.

Although, as explained above, we are not going to analyze soliton formation
and soliton in the special case of the balanced CRLH-TL with $\delta = 1$, it is worth mentioning the following.
In the case of $\delta = 1$, the dispersion relation exhibits a Dirac point, namely it is approximately
linear in the vicinity of $k=0$, i.e., $\omega \approx \pm [1+ (\beta/2)k]$ -- cf. middle panel of
Fig.~\ref{fig:dispersion}. The emergence of Dirac points is particularly interesting in the two-dimensional (2D)
setting of triangular and hexagonal lattices arising in different contexts, such as optics \cite{as},
atomic Bose-Einstein condensates \cite{carr}, and the so-called photonic graphene \cite{graph}. This has also led to an interest in this subject from a
rigorous mathematical perspective~\cite{miw}.
It is thus quite interesting that, in principle, 2D balanced CRLH-TLs may host a variety of fundamental effects,
such as conical diffraction, formation of topological defects, and even phase transitions, as in
Refs.~\cite{as,carr,graph}.

\subsection{Nonlinear analysis: the coupled NLS equations}

To describe the coupling between a RH and a LH nonlinear mode with equal group velocities, we will
use the quasi-discrete approximation, which takes into regard the inherent discreteness of the system
(see, e.g., Ref.~\cite{Remoissenet} for a review, and Refs.~\cite{lars,veldes} for relevant recent work).
Generally, this approach allows for the description of quasi-discrete envelope solitons
(usually satisfying an effective NLS model), characterized by a {\it discrete carrier} and
a slowly-varying {\it continuum envelope}. In our case, since we are interested in the description
of two different modes, we seek for a solution of Eq.~(\ref{eq:model}) in the form:
\begin{equation}
V_n=\epsilon \sum_{j=1}^{2}V_{jn}(X,T)\exp(i\theta_j) +{\rm c.c.},
\label{eq:ansatz}
\end{equation}
where ``c.c.'' denotes complex conjugate. In Eq.~(\ref{eq:ansatz}), subscripts $j=1,2$ correspond
to the LH and RH mode, $V_{jn}(X,\tau)$ are unknown (continuous) slowly-varying envelope functions
depending on the slow scales $X=\epsilon(n-v_gt)$
(where $v_g$ is the {\it common} group velocity) and $T=\epsilon^2 t$, while $\exp(i\theta_{j})$,
with $\theta_{j} = k_j n-\omega_j t$, are the (discrete) carriers of frequencies $\omega_j$ and
wavenumbers $k_j$. Finally, $\epsilon$ is a formal small parameter setting the field amplitude
and the slow scales of the envelope functions.

At this point, we should note that the field $V_n$ as expressed in Eq.~(\ref{eq:ansatz}) is, in fact,
the leading-order form of a more general ansatz employing multiple time and space scales. In this
context, use of a formal multi-scale expansion method leads to a hierarchy of equations at various powers
of $\epsilon$, which are solved up to the third-order. Here, we will present the main results and provide
further details in the Appendix A. Particularly, from the first- and second-order problems [i.e., at orders
$\mathcal{O}(\epsilon)$ (linear limit) and $\mathcal{O}(\epsilon^2)$, respectively] we derive the
dispersion relation, Eq.~(\ref{eq:linear_disp}), and the group velocity, Eq.~(\ref{eq:gvelqc}).
Finally, at the next order, $\mathcal{O}(\epsilon^3)$, we obtain the following coupled NLS equations:
\begin{eqnarray}
\!\!\!\!\!\!\!\!\!\!\!
&&i\partial_T V_1+ \frac{1}{2}D_1 \partial_{X}^2 V_1 + \left(g_{11}|V_{1}|^2+ g_{12}|V_{2}|^2\right)V_{1}=0,
\label{eq:NLS1} \\
\!\!\!\!\!\!\!\!\!\!\!
&&i\partial_T V_2+ \frac{1}{2}D_2\partial_{X}^2 V_2 + \left(g_{21}|V_{1}|^2+ g_{22}|V_{2}|^2\right)V_{2}=0,
\label{eq:NLS2}
\end{eqnarray}
where the normalized GVD coefficients $D_j$, the self-phase modulation (SPM) coefficients $g_{jj}$, and
the cross-phase modulation (CPM) coefficients $g_{j,3-j}$ (with $j=1,2$) are respectively given by:
\begin{eqnarray}
D_j&\equiv &\frac{\partial^{2}\omega_j}{\partial k_j^{2}}=
v_{g}\left[\cot k_j-\frac{\omega_j^4+3\delta^2}{\omega_j(\omega_j^4-\delta^2)}v_g\right],
\label{eq:dispcoef} \\
g_{jj}&=&\frac{\omega_j^{3}(\omega_j^{2}-\delta^{2})}{2(\omega_j^{4}-\delta^{2})}\left(3\mu-A_j\right),
\label{eq:nlcoef1} \\
g_{j,3-j}&=&\frac{\omega_j^{3}(\omega_j^{2}-\delta^{2})}{2(\omega_j^{4}-\delta^{2})}\left(6\mu-B_{3-j}\right),
\label{eq:nlcoef2}
\end{eqnarray}
and the coefficients $A_j$ and $B_{3-j}$ are defined in Appendix A.
Next, using scale transformations, we measure normalized time $T$ and densities $|V_{j}|^2$ in units
of $|D_1|^{-1}$ and $|D_1/g_{jj}|$ respectively, and cast Eqs.~(\ref{eq:NLS1})-(\ref{eq:NLS2}) in the form:
\begin{eqnarray}
&&i\partial_{T} V_1+ \frac{s}{2}\partial_{X}^2 V_1 + \left(\sigma_1|V_{1}|^2+ \lambda_1|V_{2}|^2\right)V_{1}=0,
\label{eq:MNLS1} \\
&&i\partial_{T} V_2+ \frac{d}{2}\partial_{X}^2 V_2 + \left(\lambda_2|V_{1}|^2+ \sigma_2|V_{2}|^2\right)V_{2}=0,
\label{eq:MNLS2}
\end{eqnarray}
where
\begin{eqnarray}
s&=&{\rm sign }(D_1), \quad \sigma_j={\rm sign }\:(g_{jj}),
\nonumber\\
d&=&\frac{D_2}{\left|D_1\right|},\quad
\lambda_{1}=\frac{g_{12}}{|g_{22}|},\quad
\lambda_{2}=\frac{g_{21}}{|g_{11}|}.
\label{eq:intcoeff}
\end{eqnarray}

As seen from Eqs.~(\ref{eq:MNLS1}), in the absence of coupling ($\lambda_j=0$) the evolution of either the LH mode $V_1$ or the RH mode $V_2$ is described by a single NLS equation; the latter, supports soliton solutions of the dark or the bright type, depending on the relative signs of dispersion and nonlinearity coefficients (see, e.g., Ref.~\cite{kivagr}). Particularly, the mode $V_1$ ($V_2$) supports dark solitons for $s\sigma_1<0$ ($d\sigma_2 <0$) or bright solitons for
$s\sigma_1>0$ ($d\sigma_2 >0$). These conditions, however, are modified for $\lambda_j \ne 0$ and various types of
{\it coupled} (alias vector) solitons can be found in the full version of Eqs.~(\ref{eq:MNLS1}). Below we will present these types of coupled backward- and forward-propagating solitons, belonging, respectively, to the LH and RH frequency bands.

Before proceeding with the presentation of the coupled soliton solutions we make the following comments.
First, in some cases, solitons will be found in a stationary form. However, using these stationary solutions,
one can also find travelling soliton solutions, with an additional free parameter, i.e., the velocity $C$, by means of the
following Galilean boost:
\begin{eqnarray}
V_1(X,T) &\rightarrow& V_1(X-CT,T) \nonumber \\
&\times& \exp\left\{\frac{i}{s}\left[C X+\left(\frac{C^2}{2}\right)T\right]\right\},
\label{gb1} \\
V_2(X,T) &\rightarrow& V_2(X-CT,T) \nonumber \\
&\times& \exp\left\{\frac{i}{d}\left[C X+\left(\frac{C^2}{2}\right)T\right]\right\}.
\label{gb2}
\end{eqnarray}
Second, it is
interesting to note
that, contrary to what is often
the case in the mathematically studied multi-component variants
of the NLS equation~\cite{truba}, the model of
Eqs.~(\ref{eq:MNLS1})-(\ref{eq:MNLS2}) does not necessarily
respect the condition $\lambda_1=\lambda_2$. The latter condition
ensures the existence of an underlying Hamiltonian structure
and is customary in other physical applications (such as atomic
physics~\cite{emergent}). Nevertheless, as we will see below,
this is not a necessary condition for the existence of the exact solutions
considered below.


\section{Soliton interactions in different frequency bands. Numerical results}

\subsection{Numerical procedure.}
Let us now proceed to study numerically the evolution of the coupled solitons presented
in the previous section in the framework of the fully discrete model of Eq.~(\ref{eq:model}).

In order to compare the analytical approximations with the results of numerical simulations,
we will make use of two diagnostic quantities: the first one is the evolution of the center
of mass
defined as:
\begin{equation}
X(t)=\frac{\sum_{n=-N}^{n=N} n V_n^2}{\sum_{n=-N}^{n=N} V_n^2},
\label{eq:com}
\end{equation}
and the second one, is a power-like quantity
defined as:
\begin{equation}
P(t)=\sum_{n=-N}^{n=N} V_n^2,
\label{eq:pow}
\end{equation}
with $2N+1$ being the lattice size.
The above quantities can readily be determined for each type of vector solitons that is
predicted analytically in the framework of the coupled NLS equations.

In all simulations, which have been performed by means of a fixed-step 4th-order Runge-Kutta
scheme with a time step equal to 0.01, we have fixed the value of the small parameter as
$\epsilon=0.02$, and we have used periodic boundary conditions. Use of the latter leads to
the requirement that the wavenumber $k$ of a dark soliton component must be equal to $2\pi q/p$, with
$q,p\in\mathbb{Z}$ and $q$ also being odd.

In all figures below (Figs.~\ref{fig:BBsim1}-\ref{fig:DB23sim2}), except
if stated otherwise, we
show the density plots of $V_n$, the spatial profile of $V_n$ at $t=2000$, as well as the time evolution of the center of mass $X(t)$ and the quantity
$P(t)$.

Regarding the evolution time of the simulations, we should note the following. Most of our simulations
are performed for relatively large normalized times -- typically up to $t \sim 10^7$
in some cases. However, given our time normalization, the physical unit time (set by the frequency
$f_{sh}=14.529$~GHz) is very small, namely $t_0 =(2\pi f_{sh})^{-1} \approx 11$~picoseconds (see
Sec.~II.A). Actually, since all characteristic frequencies of the system (see Eq.~(\ref{eq:freq})) are
in the microwave regime, all characteristic times are less than a nanosecond and, thus, obviously,
simulations for time $t$ even of the order of $10^9$ are extremely time-consuming. Nevertheless, our
results for normalized times up to $t=10^7$ (corresponding to a physical time of the
order of a tenth of millisecond), demonstrate a good agreement with our analytical predictions in
suitable cases (see below). Furthermore, the results of such long simulations can also be used as a
reliable indication of the solitons' robustness. Hence, in the case where
the solitary waves are found to be very robust, we expect that they would
survive for the longer time scales that would render them experimentally
observable.

\subsection{Bright-bright solitons in bands II and IV.}

First, we consider the interaction between a backward propagating soliton, with a frequency lying in band II, and a forward propagating soliton, with a frequency lying in band IV. In this case, $s=-1$ (cf. Fig.~\ref{fig:gvel}), while the other dispersion and nonlinearity coefficients are shown in Fig.~\ref{fig:INTBDcoeff} as functions of the normalized frequency $f/f_{\rm sh}$ (for $\delta=1.0954$). It is observed that $\sigma_1=\sigma_2=-1$, and also $\lambda_1 \approx \lambda_2\equiv\lambda $, while the dispersion coefficient $d$ takes values $d \lesssim -0.25$. Thus, to a first approximation, the system of Eqs.~(\ref{eq:MNLS1})-(\ref{eq:MNLS2}) takes the form:
\begin{eqnarray}
&&i\partial_{T} V_1- \frac{1}{2}\partial_{X}^2 V_1 + \left(\lambda |V_{2}|^2-|V_{1}|^2\right)V_{1}=0,
\label{eq:MNLSMAN1}\\
&&i\partial_{T} V_2+ \frac{d}{2}\partial_{X}^2 V_2 + \left(\lambda |V_{1}|^2-|V_{2}|^2\right)V_{2}=0.
\label{eq:MNLSMAN2}
\end{eqnarray}
where $\lambda<0$ and $d<0$, as can be seen in the top panel of Fig.~\ref{fig:INTBDcoeff}.
The above system is generally non-integrable and soliton solutions can not be found in an explicit analytical form.
However, there exists a specific frequency value, namely $f/f_{\rm sh}=0.9654$ where the dispersion coefficient $d$ take the value $d\simeq-1$ and the nonlinearity coefficients $\lambda_{1,2}$ take the values $\lambda_1=\lambda_2=-1.7$ (see the intersection point of the relevant curves depicted by a star in the top panel of Fig.~\ref{fig:INTBDcoeff}). This case corresponds
to a (common for both components) group velocity $v_g=0.1288$, which occurs when the (normalized) carrier frequencies for the modes $V_1$ and $V_2$ take, respectively, the values $f_1/f_{\rm sh}=0.9654$ (as mentioned above) and $f_2/f_{\rm sh}=1.3653$. Then, symmetric bright-bright
standing soliton solutions can be found in the following form (see, e.g., Ref.~\cite{yang}):
\begin{eqnarray}
V_1=V_2=\sqrt{\frac{2\ell}{1-\lambda}}\sech(\sqrt{2\ell}X)\exp(-i\ell T),
\label{eq:sol1}
\end{eqnarray}
%
where $\ell$ is an arbitrary parameter.
Using the above expressions, we can now approximate the unknown voltage $V_n(t)$ in Eq.~(\ref{eq:model}), in terms of the original coordinates $n$ and $t$, as follows:
\begin{figure}[tbp]
\includegraphics[scale=0.4]{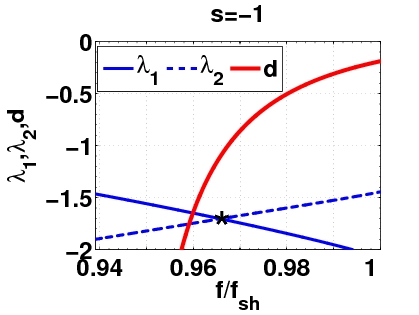}
\includegraphics[scale=0.4]{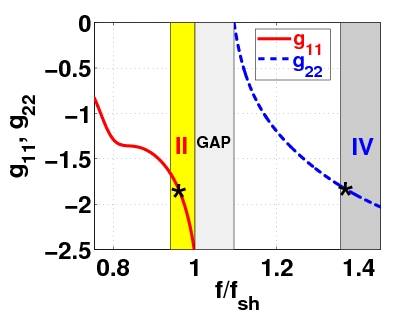}
\caption{(Color online) Parameters for soliton interactions in bands II and IV. Top panel: the dependence of the parameters $\lambda_1$ [thin solid (blue) line], $\lambda_2$ [dashed (blue) line] and $d$ [bold solid (red) line] on the normalized frequency $f/f_{\rm sh}$. Bottom panel: the nonlinearity coefficients $g_{11}$ [solid (red) line] and $g_{22}$ [dashed (blue) line] as functions of $f/f_{\rm sh}$. The parameter $\delta$ takes the value $\delta=1.0954$. Stars (in black) in both panels corresponding to $f/f_{\rm sh}=0.9654$, for which $d=\sigma_{1,2}=-1$ and $\lambda_1=\lambda_2=\lambda=-1.7$. These values are used for the simulations shown in
Figs.~\ref{fig:BBsim1} and \ref{fig:BBsim2} below.}
\label{fig:INTBDcoeff}
\end{figure}
\begin{eqnarray}
V_{n}(t) &\approx&
V_{0}
[R_1(n,t) \cos(k_{1} n-\Omega_{1}t)\nonumber\\
&+&R_2(n,t) \cos(k_{2} n-\Omega_{2}t),
\label{eq:bright1}
\end{eqnarray}
where functions $R_1$ and $R_2$ have the following form:
\begin{eqnarray}
&&R_1=\sech[\sqrt{2\ell}\epsilon (n-v_g t)],\\
\label{eq:bs1}
&&R_2=\sqrt{\left|\frac{g_{11}}{g_{22}}\right|}\sech[\sqrt{2\ell}\epsilon (n-v_g t)],
\label{eq:bs2}
\end{eqnarray}
while the solution amplitude $V_0$ and the frequencies $\Omega_{j}$ ($j=1,2$) are given by:
\begin{eqnarray}
&&V_{0}=2\epsilon\sqrt{\left|\frac{D_1}{g_{11}}\right|\frac{2\ell}{1-\lambda}},
\label{v01} \\
&&\Omega_{j}=\omega_j+ \epsilon^2 \ell |D_1|,
\end{eqnarray}
with $\omega_j \equiv f_j/f_{\rm sh}$.
Now, substituting Eq.~(\ref{eq:bright1}) into Eqs.~(\ref{eq:com}) and (\ref{eq:pow}) and supposing that $\epsilon$ is small enough,
we obtain for our diagnostic quantities:
\begin{eqnarray}
X(t)&=&v_g t, \\
\label{eq:XBB}
P(t)&=&
\frac{V_0^2}{\epsilon\sqrt{2\ell}}\left(1+\left|\frac{g_{11}}{g_{22}}\right|\right).
\label{eq:PBB}
\end{eqnarray}

\begin{figure}[tbp]
    \begin{tabular}{cc}
    \includegraphics[width=3.85cm]{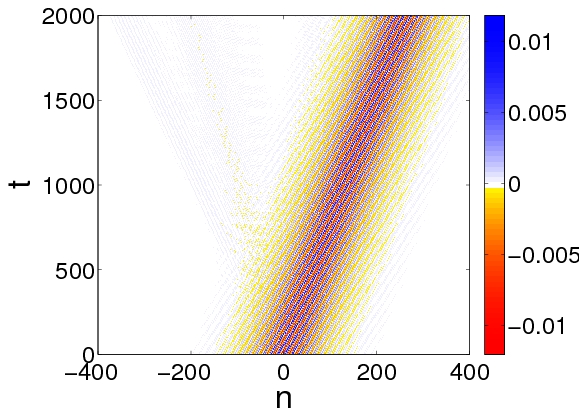} &
    \includegraphics[width=3.85cm]{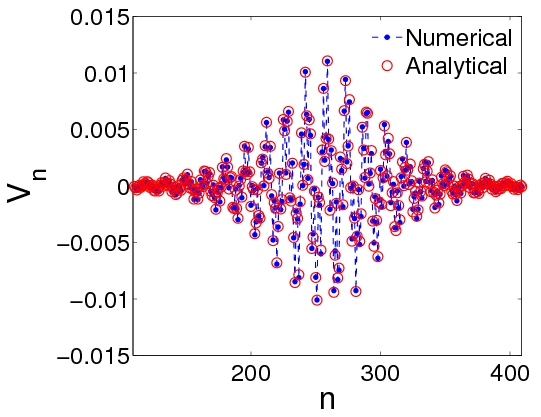} \\
    \includegraphics[width=3.85cm]{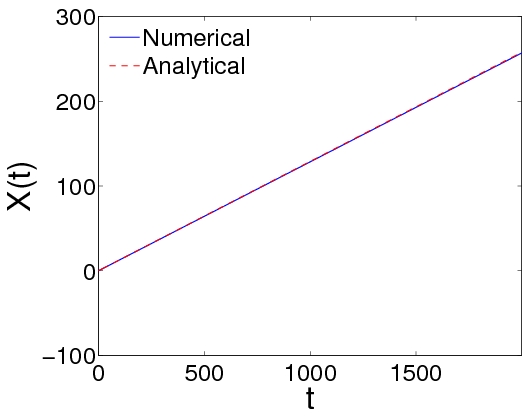} &
    \includegraphics[width=3.85cm]{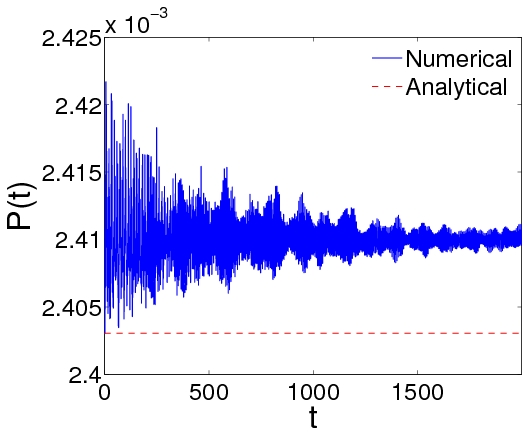} \\
    \end{tabular}
\caption{(Color online) Bright-bright solitons in regions II-IV. Top left: density plot of the space-time evolution of $V_n$ obtained numerically. The top right panel compares the analytical and numerical profiles of $V_n$ at $t=2000$. The bottom panels show the time evolution of the center of mass (left) and the power diagnostic (right). The parameters used are $f_1=0.96545$ and $f_2=1.36535$, which gives $k_1=-0.4061$ and $k_2=1.8576$, i.e. a bright-bright soliton in bands II and IV (this
particular choice corresponds to the points depicted by stars in Fig.~\ref{fig:INTBDcoeff}). The difference in the powers can be attributed to the approximate
nature of our solution.}
\label{fig:BBsim1}
\end{figure}

\begin{figure}[tbp]
    \begin{tabular}{cc}
    \includegraphics[width=3.85cm]{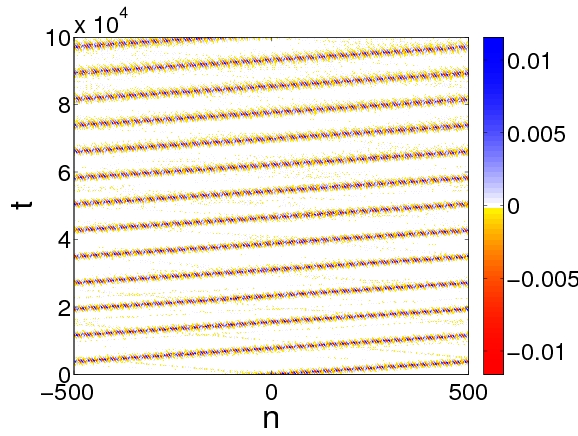} &
    \includegraphics[width=3.85cm]{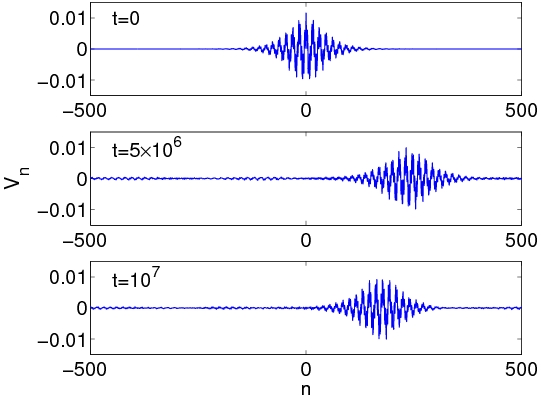} \\
    \includegraphics[width=3.85cm]{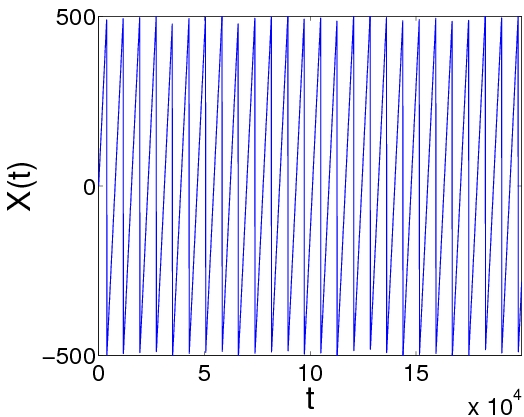} &
    \includegraphics[width=3.85cm]{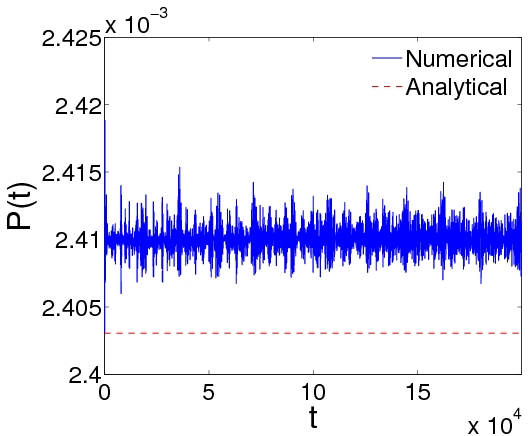} \\
    \end{tabular}
\caption{(Color online) The bright-bright soliton of Fig.~\ref{fig:BBsim1} is evolved until $t=10^5$.
All the panels are similar to that of Fig.~\ref{fig:BBsim1} except for the top right. In this panel, snapshots of the soliton at $t=2\times10^6$ and $t=10^7$ are compared to the initial condition of the simulation in order to examine its robustness under a very long evolution time.}
\label{fig:BBsim2}
\end{figure}

In Figs.~\ref{fig:BBsim1} and \ref{fig:BBsim2} we show the outcome of the simulations for short and long times, respectively, of a bright-bright soliton with $\ell=1$ and $N=500$.
The parameters used are $f_1=0.96545$ and $f_2=1.36535$, which gives $k_1=-0.4061$ and $k_2=1.8576$, i.e., a bright-bright soliton in bands II and IV.
In Fig.~\ref{fig:BBsim1}, it is  evident
that the agreement between analytical and numerical results pertaining to the soliton profile, and the evolution of the center of mass and power diagnostics, is very good.
In the case shown in Fig.~\ref{fig:BBsim2}, we have performed a very long simulation, up to normalized times $t=10^7$.
It is clear that that the initial pulse does not spread out, which indicates
the soliton robustness:
the top panels of the figure -- and particularly the snapshots of the pulse profile at $t = 10^7$ -- clearly show that the soliton persists as a stable object up to the end of
this long simulation time.
We note in passing that
a fragment of the soliton is backscattered when
the soliton starts its motion at $t=0$ (due to the approximate
nature of our analytical solution profile). Notice that despite this
emission and the subsequent interaction of the fragment with the ``distilled''
solitary wave, the coherent structure remains robust and preserves its
characteristics throughout the evolution thereafter.

\subsection{Bright-dark solitons in bands I and IV}

\begin{figure}[tbp]
\includegraphics[scale=0.4]{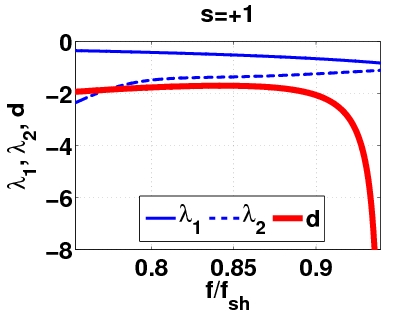}
\includegraphics[scale=0.4]{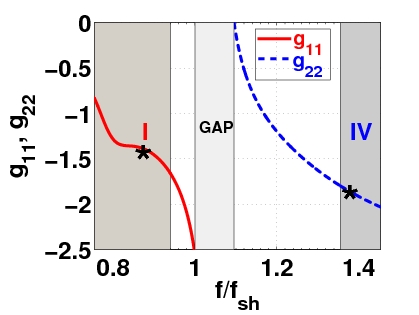}
\caption{(Color online) Same as Fig.~\ref{fig:INTBDcoeff}, but for soliton
interactions in bands I and IV. Stars depict parameter values used for the simulations
shown in Figs.~\ref{fig:DB14sim1} and \ref{fig:DB14sim2} below.}
\label{fig:INTADcoeff}
\end{figure}

Next, we consider the interaction between a backward propagating soliton, with a frequency lying in band I, and a forward propagating soliton, with a frequency lying in band IV. In Fig.~\ref{fig:INTADcoeff} we show the dependence of the parameters $\lambda_1$, $\lambda_2$ and $d$ (top panel), as well as of the nonlinearity coefficients (bottom panel), on the normalized frequency $f/f_{\rm sh}$ (for $\delta=1.0954$). In this case, $s=+1$ (cf. Fig.~\ref{fig:gvel}), while $\sigma_1=\sigma_2=-1$ and, thus, Eqs.~(\ref{eq:MNLS1})-(\ref{eq:MNLS2}) are reduced to the form:
\begin{eqnarray}
&&i\partial_{T} V_1+ \frac{1}{2}\partial_{X}^2 V_1 + \left(\lambda_1|V_{2}|^2-|V_{1}|^2 \right)V_{1}=0,
\label{eq:MNLSAD1}\\
&&i\partial_{T} V_2+ \frac{d}{2}\partial_{X}^2 V_2 + \left(\lambda_2|V_{1}|^2-|V_{2}|^2 \right)V_{2}=0,
\label{eq:MNLSAD2}
\end{eqnarray}
where $\lambda_{1,2}<0$ and $d<0$, as can be seen in the top panel of Fig.~\ref{fig:INTADcoeff}.
The above equations are no longer of the Manakov type and, thus, generally, they are not completely integrable. Nevertheless, standing wave solutions can still be found in the form of coupled bright-dark solitons, with the frequency of the bright (dark) soliton component being in the LH (RH) frequency band. Therefore, here we have a case of coupled solitons, a backward-propagating bright soliton and a forward-propagating dark soliton, whose exact analytical form is:
\begin{eqnarray}
\!\!\!\!
V_1(X,T)&=&\sqrt{\frac{\nu_2(1+d\lambda_1)}{|d+\lambda_2|)}} {\rm sech}(b X)\exp(-i\nu_1 T),
\label{eq:bs}\\
\!\!\!\!
V_2(X,T)&=&\sqrt{\nu_2} \tanh(b X)\exp(-i\nu_2 T),
\label{eq:ds}
\end{eqnarray}
where the soliton amplitude parameters $\nu_{1,2}$ and the inverse width $b$ are connected via the following equations:
\begin{eqnarray}
&&\nu_1=-\frac{\nu_2}{2(d +\lambda_2)}[1+(2d+\lambda_2)\lambda_1],
\label{eq:n1}\\
&&b^2=\frac{\nu_2}{d+ \lambda_2}(1-\lambda_1\lambda_2),
\label{eq:bb}
\end{eqnarray}
with $1-\lambda_1\lambda_2<0$ in the considered LH frequency band (notice that $d+ \lambda_2<0$ as
well). It is thus clear that the above solutions are characterized by one free parameter.

\begin{figure}[tbp]
    \begin{tabular}{cc}
    \includegraphics[width=3.85cm]{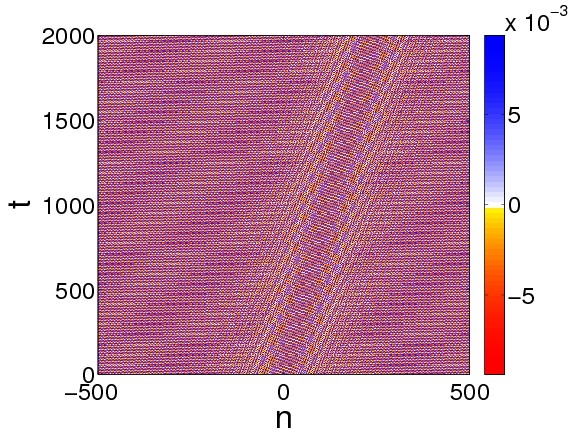} &
    \includegraphics[width=3.85cm]{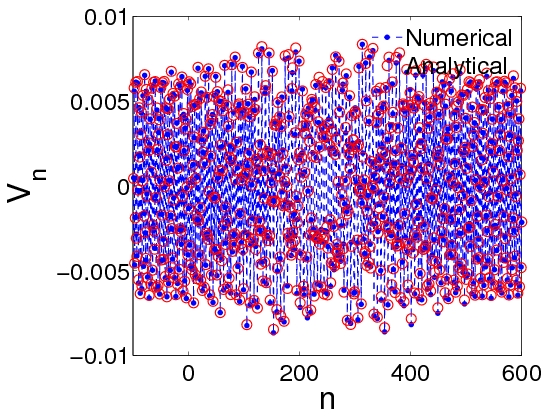} \\
    \includegraphics[width=3.85cm]{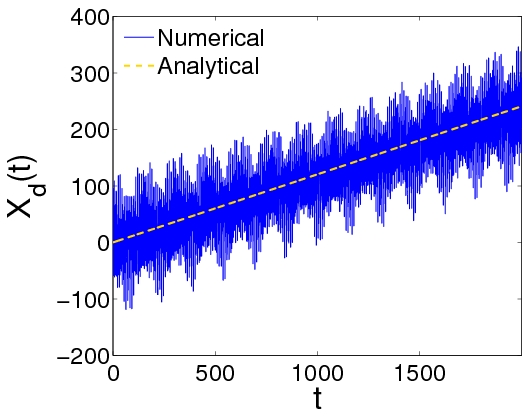} &
    \includegraphics[width=3.85cm]{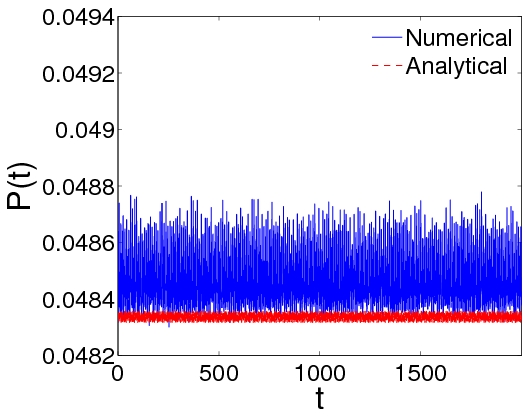} \\
    \end{tabular}
\caption{(Color online) Bright-dark solitons in regions I-IV. Top left: density plot of the time evolution of $V_n$ obtained numerically. The top right panel compares the analytical and numerical profiles of $V_n$ at $t=2000$. The bottom panels show the time evolution of the center of mass (left) and the power diagnostics (right). The parameters used are $f_1=0.8831$ and $k_2=5\pi/8\approx1.9625$, which gives $k_1=-1.0404$ and $f_2=1.3748$
(see corresponding points depicted by stars in Fig.~\ref{fig:INTADcoeff}),
i.e. a bright-dark soliton in the I and IV bands.}
\label{fig:DB14sim1}
\end{figure}
\begin{figure}[h]
    \begin{tabular}{cc}
    \includegraphics[width=3.85cm]{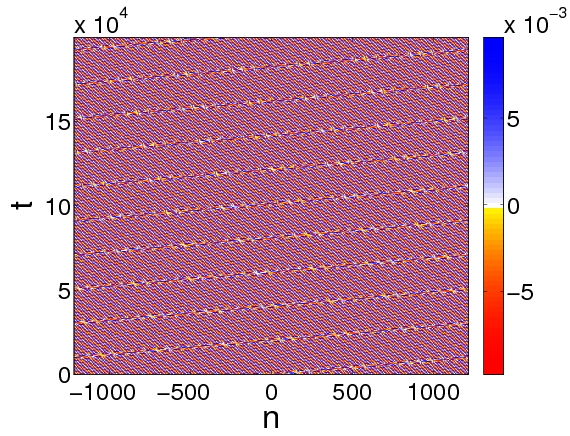} &
    \includegraphics[width=3.85cm]{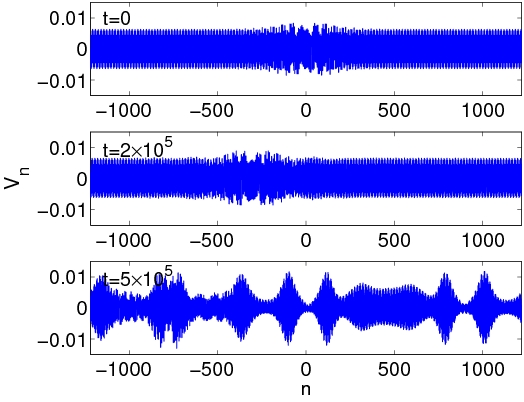} \\
    \includegraphics[width=3.85cm]{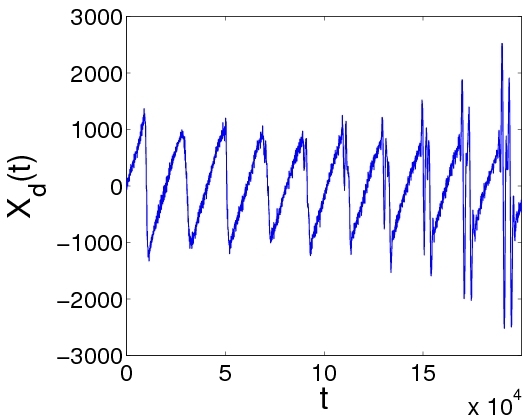} &
    \includegraphics[width=3.85cm]{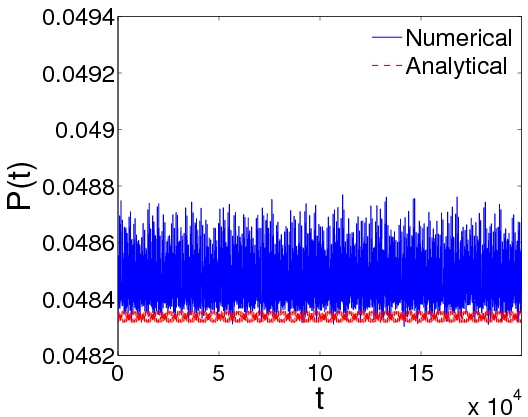} \\
    \end{tabular}
\caption{(Color online) The bright-dark soliton of Fig.~\ref{fig:BBsim1} is evolved until $t=2\times10^5$. All the panels are similar to that of Fig.~\ref{fig:DB14sim1} except for the top right. In this panel, snapshots of the soliton at $t=2\times10^5$ and $t=5\times10^5$ are compared to the initial condition of the simulation. Notice that the center of mass is not bounded into $[-N,N]$; this can be caused by the soliton splitting.}
\label{fig:DB14sim2}
\end{figure}

Employing the solutions (\ref{eq:bs})-(\ref{eq:ds}), we can again approximate a solution of Eq.~(\ref{eq:model}) for the voltage $V_n(t)$, in terms of the original coordinates $n$ and $t$, as follows:
\begin{eqnarray}
V_{n}(t)& \approx & V_{0}[\Psi_1(n,t) \cos(k_{1} n-\Omega_{1}t)\nonumber\\
&+&\Psi_2(n,t) \cos(k_{2} n-\Omega_{2}t)],
\label{eq:brightdark1}
\end{eqnarray}
where
\begin{eqnarray}
\!\!\!\!\!\!\!\!
&&\Psi_1=\Psi_{1,0}\sech[\epsilon b(n-v_g t)],\ \Psi_{1,0}=\sqrt{\frac{1+d\lambda_1}{|d+\lambda_2|}},
\label{eq:dbsol1}\\
\!\!\!\!\!\!\!\!
&&\Psi_2=\Psi_{2,0}\tanh[\epsilon b(n-v_g t)],\ \Psi_{2,0}=\sqrt{\left|\frac{g_{11}}{g_{22}}\right|}.
\label{eq:dbsol2}
\end{eqnarray}
In this case, the solution amplitude $V_0$ and the frequencies $\Omega_{j}$ ($j=1,2$) are given by:
\begin{eqnarray}
V_{0}&=&2\epsilon\sqrt{\nu_2\left|\frac{D_1}{g_{11}}\right|},
\label{v02} \\
\Omega_{j}&=&\omega_j+ \epsilon^2\nu_j|D_1|.
\end{eqnarray}

In order to get an expression for the center of mass similar to that of the bright-bright soliton (\ref{eq:XBB}), we must define it as:

\begin{eqnarray}
\!\!\!\!\!\!\!\!
X_d(t)=\frac{\ F(t)X(t)}{\Psi_{1,0}^2-\Psi_{2,0}^2}-\frac{\epsilon b N}{2}\Psi_{2,0}^2\cot(k_2)\sin(2\Omega_2 t),
\label{eq:comBD}
\end{eqnarray}
where
\begin{equation}
\!\!\!\!\!\!\!\!
F(t)={\Psi_{1,0}^2+\Psi_{2,0}^2(\epsilon b N-1)+\frac{\epsilon b}{2} \Psi_{2,0}^2[1-\cos(2\Omega_2 t)]}.
\label{eq:comBD_F}
\end{equation}

Substituting Eq.~(\ref{eq:brightdark1}) into  Eqs.~(\ref{eq:comBD}) and (\ref{eq:pow}) we can once again obtain relevant expressions (provided that $\epsilon$ is small enough) for the center of mass and power:
\begin{equation}
X_d(t)=v_g t,
\label{eq:XBD}
\end{equation}
\begin{equation}
P_d(t)=\frac{V_0^2}{\epsilon b}F(t).
\label{eq:PBD}
\end{equation}

Figures~\ref{fig:DB14sim1} and \ref{fig:DB14sim2} show the evolution of a bright-dark soliton
(and its characteristics) in bands I and IV with $\nu_2=1$ and $N=1220$. The parameters used are $f_1=0.8831$ and $k_2=5\pi/8\approx1.9625$, which give $k_1=-1.0404$ and $f_2=1.3748$. In this  case, it is clear that although bright-dark solitons do exist, the agreement between analytical and numerical results becomes worse over time. Also, as shown in the top right panel of Fig.~\ref{fig:DB14sim2}, the pulse profile indicates that the bright-dark soliton is not a robust
object. In particular, at time $t=5 \times 10^5$, it is clear that the
configuration has dramatically changed its character.

\subsection{Dark-bright solitons in bands I-III and II-III }
\begin{figure}[tbp]
\includegraphics[scale=0.4]{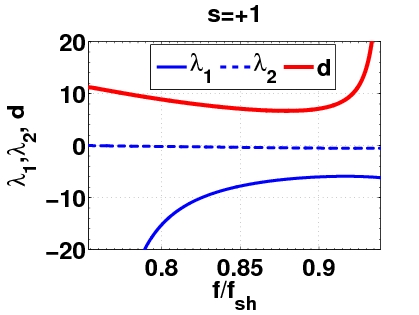}
\includegraphics[scale=0.4]{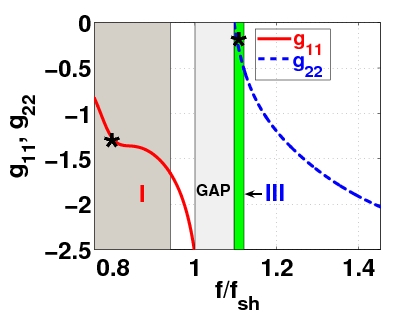}
\caption{(Color online) Same as Fig.~\ref{fig:INTBDcoeff}, but for soliton interactions
in bands I and III. Stars depict parameter values used for the simulations
shown in Figs.~\ref{fig:DB13sim1} and \ref{fig:DB13sim2} below.}
\label{fig:INTACcoeff}
\end{figure}
\begin{figure}[htbp]
\includegraphics[scale=0.4]{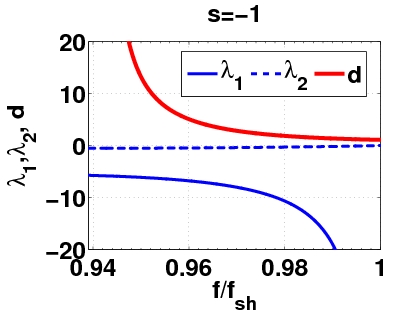}
\includegraphics[scale=0.4]{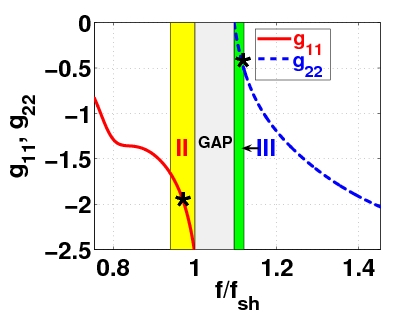}
\caption{(Color online) Same as Fig.~\ref{fig:INTBDcoeff},
but for soliton interactions in bands II and III. Stars depict parameter values used
for the simulations shown in Figs.~\ref{fig:DB23sim1} and \ref{fig:DB23sim2} below.}
\label{fig:INTBCcoeff}
\end{figure}

Finally, we consider the cases of coupled solitons in bands I and III, and also in
bands II and III. In both cases,
as is observed in the top panels of Figs.~\ref{fig:INTACcoeff} (bands I-III) and \ref{fig:INTBCcoeff}
(bands II-III), we have that
the parameter $\lambda_2\ll1$. Now, the NLS equations for $V_1$ and $V_2$ take the following form:
\begin{eqnarray}
&&i\partial_{T} V_1+ \frac{s}{2}\partial_{X}^2 V_1 + \left(\lambda_1|V_{2}|^2-|V_{1}|^2 \right)V_{1}=0,
\label{eq:MNLSAC1}\\
&&i\partial_{T} V_1+ \frac{d}{2}\partial_{X}^2 V_1 + \left(\lambda_2|V_{1}|^2-|V_{2}|^2 \right)V_{2}=0,
\label{eq:MNLSAC2}
\end{eqnarray}
where $s=+1$ ($s=-1$) corresponds to solitons in bands I and III (II and III). Note that, in this case, $\sigma_1=\sigma_2=-1$, while $d>0$ and $\lambda_{1,2}<0$, as shown in Figs.~\ref{fig:INTACcoeff} and \ref{fig:INTBCcoeff}.
As in the previous case, the  equations  ~(\ref{eq:MNLSAC1})-(\ref{eq:MNLSAC2}), generally,  are not completely integrable. Nevertheless, standing wave solutions can still be found in the form of coupled dark-bright solitons, with the frequency of the dark (bright) soliton component being in the LH (RH) frequency band; therefore, here we have a case of a backward-propagating dark soliton, coupled with
a forward-propagating bright soliton, whose exact analytical forms are:
\begin{eqnarray}
\!\!\!\!\!
V_1(X,T)&=&\sqrt{\eta_1} {\rm tanh(BX)}\exp(-i\eta_1 T),\\
\label{eq:DS13}
\!\!\!\!\!
V_2(X,T)&=&\sqrt{\frac{\eta_1(d+s\lambda_2)}{|d\lambda_1+s|}} {\rm sech(BX)}\exp(-i\eta_2 T),
\label{eq:BS13}
\end{eqnarray}
where the soliton amplitude parameters $\eta_{1,2}$ and the inverse width $B$ are connected via the following equations:
\begin{eqnarray}
&&\eta_2=-\frac{\eta_1}{2(d\lambda_1 +s)}[d+(d\lambda_1+2s)\lambda_2],
\label{eq:eta2}\\
&&B^2=\frac{\eta_1}{d\lambda_1+ s }(1-\lambda_1\lambda_2).
\label{eq:BB}
\end{eqnarray}
\begin{figure}[tbp]
    \begin{tabular}{cc}
    \includegraphics[width=3.85cm]{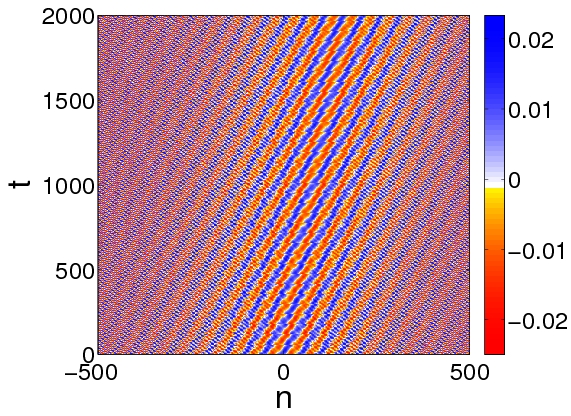} &
    \includegraphics[width=3.85cm]{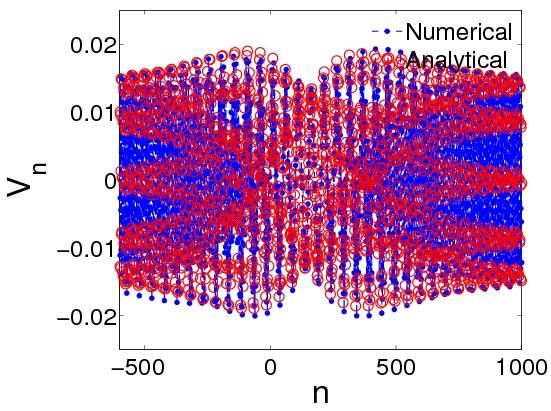} \\
    \includegraphics[width=3.85cm]{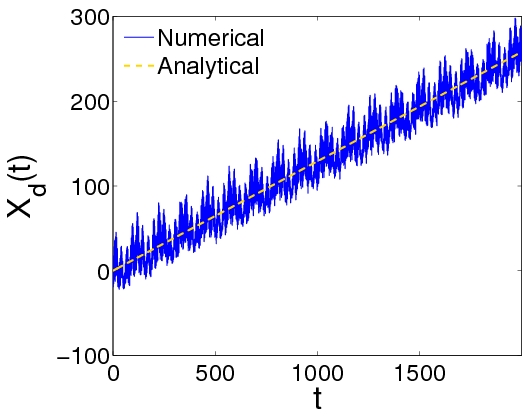} &
    \includegraphics[width=3.85cm]{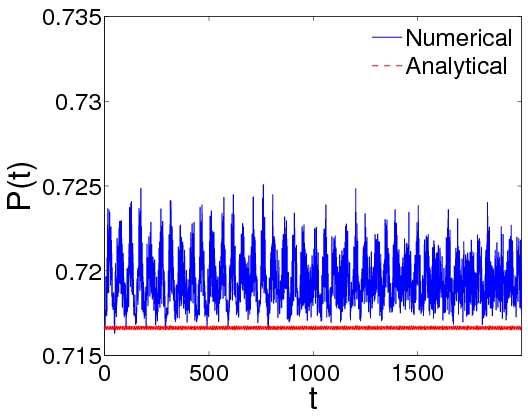} \\
    \end{tabular}
\caption{(Color online) Dark-bright solitons in regions I-III. Top left: density plot of the time
evolution of $V_n$ obtained numerically. The top right panel compares the analytical and numerical
profiles of $V_n$ at $t=2000$. The bottom panels show the time evolution of the center of mass (left)
and the width diagnostic (right). Parameters used are $k_1=-6\pi/5\approx-1.884$ and
$f_2=1.1002$, which gives $f_1=0.8003$ and $k_2=0.1232$ (cf. points depicted by stars
in Fig.~\ref{fig:INTACcoeff}), i.e. a dark-bright soliton in bands I and III.}
\label{fig:DB13sim1}
\end{figure}
\begin{figure}[tbp]
    \begin{tabular}{cc}
    \includegraphics[width=3.85cm]{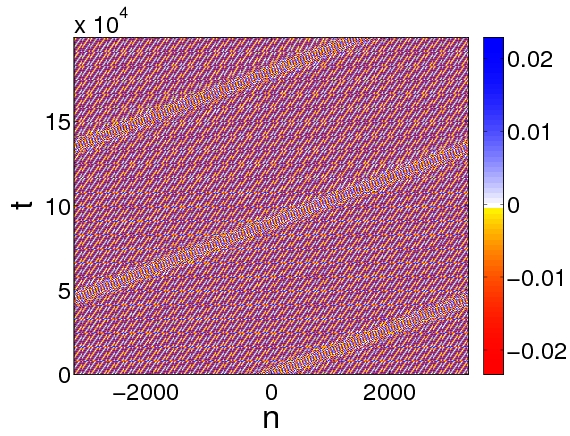} &
    \includegraphics[width=3.85cm]{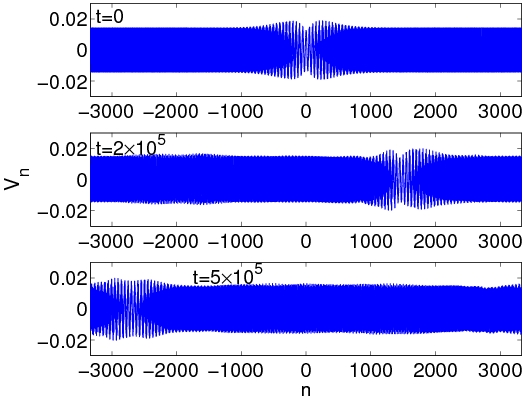} \\
    \includegraphics[width=3.85cm]{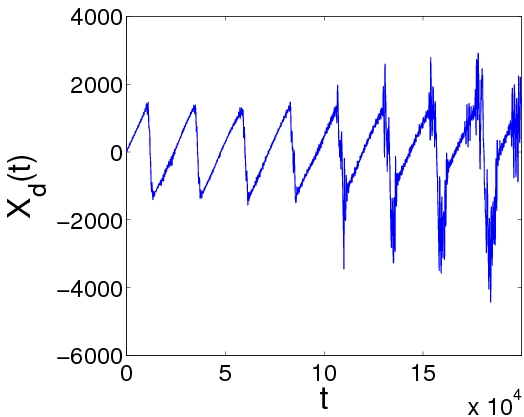} &
    \includegraphics[width=3.85cm]{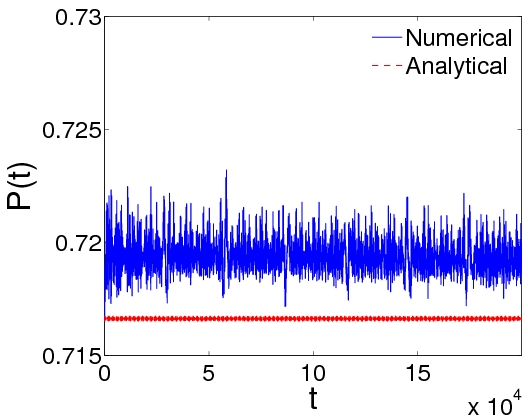} \\
    \end{tabular}
\caption{(Color online) The dark-bright soliton of Fig.~\ref{fig:DB13sim1} is evolved until $t=2\times10^5$.
All the panels are similar to that of Fig.~\ref{fig:DB13sim1} except for the top right. In this panel,
snapshots of the soliton at $t=2\times10^5$ and $t=5\times10^5$ are compared to the initial condition
of the simulation in order to examine its robustness under a very lengthy time evolution.}
\label{fig:DB13sim2}
\end{figure}
Following our previous considerations, we may again use the above solutions and approximate
the voltage $V_n(t)$ in Eq.~(\ref{eq:model}), in terms of the original coordinates, as follows:
\begin{eqnarray}
V_{n}(t)&=& V_{0}[\Phi_1(n,t) \cos(k_{1} n-\Omega_{1}t)\nonumber\\
&+&\Phi_2(n,t) \cos(k_{2} n-\Omega_{2}t)],
\label{eq:darkbright1}
\end{eqnarray}
where functions $\Phi_1$ and $\Phi_2$ are given by:
\begin{eqnarray}
\Phi_1&=&\tanh[\epsilon B(n-v_g t)],
\label{eq:dbsol131}\\
\Phi_2&=& \Phi_{2,0}\sech[\epsilon B(n-v_g t)],\nonumber\\
\Phi_{2,0}&=&\sqrt{\frac{(d+s\lambda_2)}{|d\lambda_1+s|}\left|\frac{g_{11}}{g_{22}}\right|},
\label{eq:dbsol132}
\end{eqnarray}
while the rest of the soliton parameters are:
\begin{eqnarray}
&&V_{0}=2\epsilon\sqrt{\eta_1\left|\frac{D_1}{g_{11}}\right|},
\label{v03} \\
&&\Omega_{j}=\omega_j+ \epsilon^2\eta_j|D_1|.
\label{omm}
\end{eqnarray}

\begin{figure}[tbp]
    \begin{tabular}{cc}
    \includegraphics[width=3.85cm]{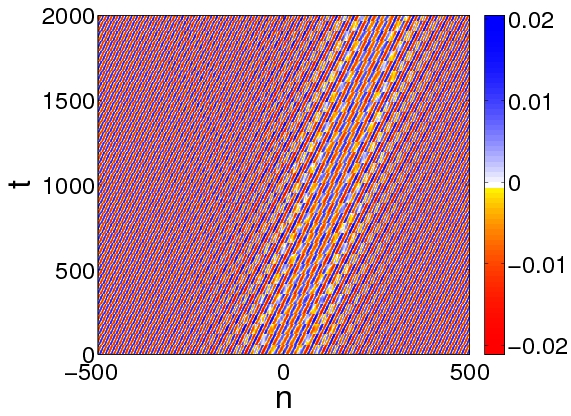} &
    \includegraphics[width=3.85cm]{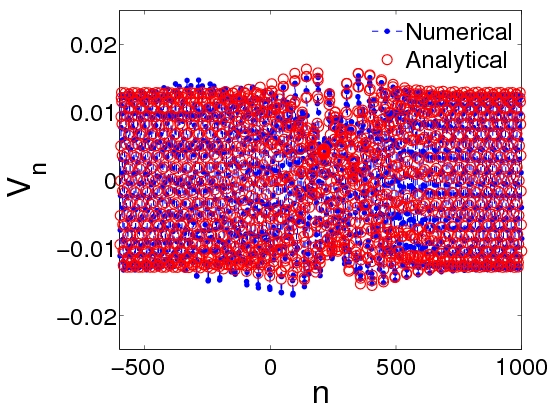} \\
    \includegraphics[width=3.85cm]{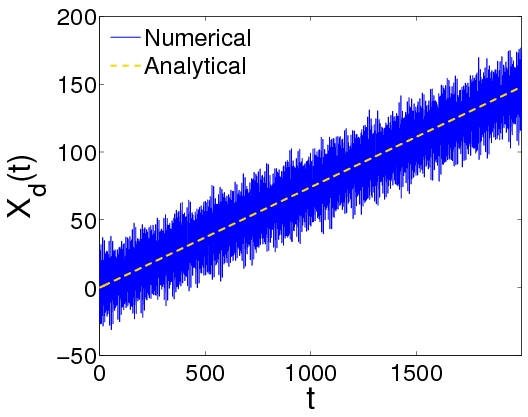} &
    \includegraphics[width=3.85cm]{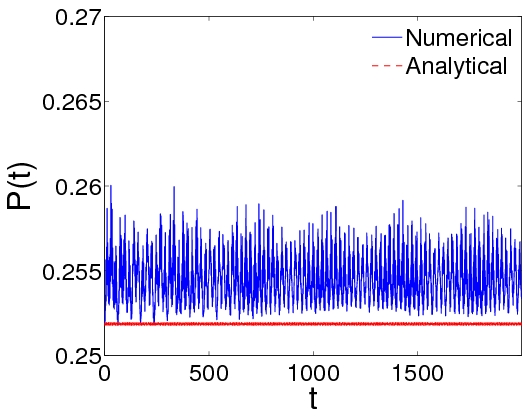} \\
    \end{tabular}
\caption{(Color online) Dark-bright solitons in regions II-III. Top left: density plot of the time evolution
of $V_n$ obtained numerically. The top right panel compares the analytical and numerical profiles of $V_n$
at $t=2000$. The bottom panels show the time evolution of the center of mass (left) and the width diagnostic (right).
Parameters used are $k_1=-3\pi/23\approx-0.4095$ and $f_2=1.1162$, which gives $f_1=0.965$ and $k_2=0.2758$
(cf. stars in Fig.~\ref{fig:INTBCcoeff}), i.e. a dark-bright soliton in bands II and III.}
\label{fig:DB23sim1}
\end{figure}

\begin{figure}[tbp]
    \begin{tabular}{cc}
    \includegraphics[width=3.85cm]{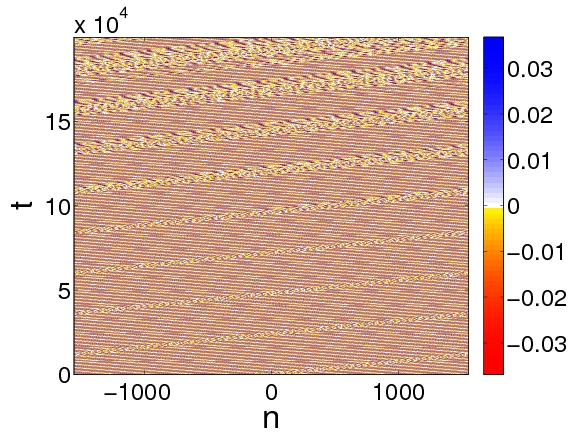} &
    \includegraphics[width=3.85cm]{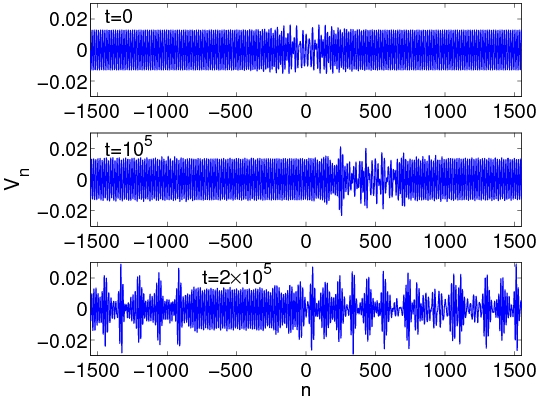} \\
    \includegraphics[width=3.85cm]{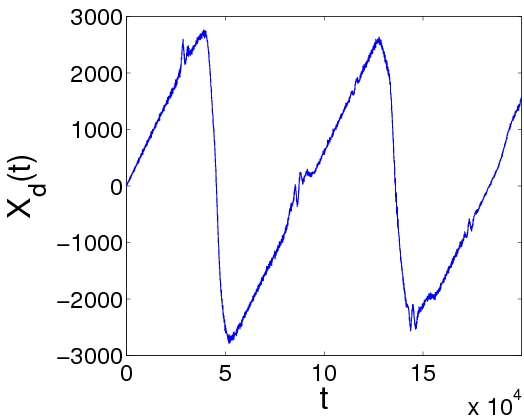} &
    \includegraphics[width=3.85cm]{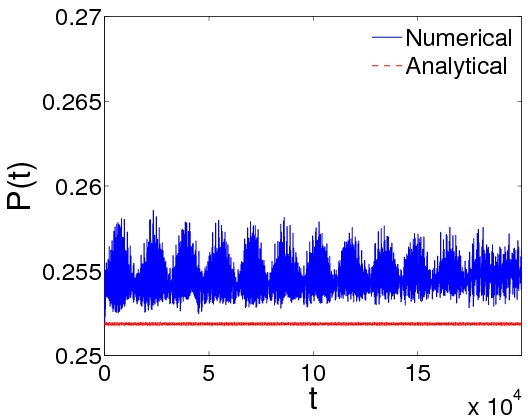} \\
    \end{tabular}
\caption{(Color online) The dark-bright soliton of Fig.~\ref{fig:DB23sim1} is evolved until $t=2\times10^5$.
All the panels are similar to that of Fig. \ref{fig:DB23sim1} except for the top right. In this panel,
snapshots of the soliton at $t=10^5$ and $t=2\times10^5$ are compared to the initial condition of the simulation.
Notice that, as in Fig. \ref{fig:DB14sim2}, the center of mass is not bounded into $[-N,N]$; as in that case,
the soliton splits at long evolution times.}
\label{fig:DB23sim2}
\end{figure}

%

Figures~\ref{fig:DB13sim1} and \ref{fig:DB13sim2} show the outcome of the simulations for
a dark-bright soliton in bands I and III (with $\eta_1=3$ and $N=3333$), while
Figs.~\ref{fig:DB23sim1} and \ref{fig:DB23sim2} correspond to a dark-bright soliton
in bands II and III (with $\eta_1=3$ and $N=1553$). The parameters used are $k_1=-6\pi/5\approx-1.884$
and $f_2=1.1002$, which gives $f_1=0.8003$ and $k_2=0.1232$, in bands I and III, and $k_1=-3\pi/23\approx-0.4095$ and
$f_2=1.1162$, which gives $f_1=0.965$ and $k_2=0.2758$, in
bands II and III, respectively.
In the latter case, the relatively large values of
the number of particles and of
$\eta_1$ used
are motivated by the necessity of a vanishing tail for the bright component at the edges of the
lattice. As seen in this set of figures, dark-bright solitons in bands II-III and I-III do
exist, as predicted in theory. Furthermore, it is observed that the former are less robust
than the latter, as seen both from their stronger deformation and the fact that they ``lose''
their solitary wave character earlier.
This can be observed, e.g., in the strong fluctuations in the evolution of the
soliton center in the bottom left panel of Fig.~\ref{fig:DB23sim2}, or perhaps
most notably in the substantial modification of the wave profile upon
long propagation in the top right panel of the same figure.
On the other hand, the dark-bright solitons of bands I-III seem to essentially
preserve their structure even in the long evolution of
Fig.~\ref{fig:DB13sim2}.

\section{Conclusions}

In conclusion, we have used both analytical and numerical techniques to study the
existence, stability and dynamics of coupled backward- and forward-propagating solitons in
a composite right/left-handed (CRLH) nonlinear transmission line (TL). The considered form of
the TL was a quite generic one, finding applications to the modelling of a wide range of
LH systems and devices, with ``parasitic'' RH behavior, such as resonators, antennas, directional couplers, among others~\cite{review2,review1,review3,Caloz1}.

Our analysis started with the derivation of a nonlinear lattice equation governing the voltage
across the fundamental (unit cell) element of the transmission line.
In the linear regime, we derived the dispersion relation for small-amplitude linear plane waves
and showed that they may either propagate in a right-handed (RH) high-frequency region, or in a
left-handed (LH) low-frequency region. We also identified frequency bands where RH- and LH-modes
can propagate with the same group velocity.

Using the above result, we then investigated the possibility of nonlinearity-assisted coupling between
LH- and RH-modes. This way, in order to analytically treat the nonlinear lattice equation, we used the
so-called quasi-discrete approximation. The latter is a variant of the multi-scale perturbation method,
which takes into regard the discreteness of the system by considering the carrier (envelope) of the
wave as a discrete (continuum) function of space. Employing this approach, we derived,
in the small-amplitude approximation and for certain space- and time-scales, a system of
two coupled nonlinear Schr\"{o}dinger (NLS) equations for the unknown voltage envelope functions.
This system was then used to predict the existence of coupled backward- and forward-propagating
solitons, of the bright-bright, bright-dark and dark-bright type, respectively.

The above existence results, as well as the propagation properties and the
potential robustness of these vector
solitons, were then investigated for each of the possible
scenarios. This was done
by means of direct numerical simulations of the full CRLH-TL nonlinear
lattice model,
using as initial
conditions the analytical forms of solitons predicted by the perturbation theory. In the simulations,
apart from the evolution of the shape, we also studied the evolution of the center of mass
and a power-like quantity of the various solitons. Our numerical results have confirmed the existence of the
various types of solitons predicted analytically, but have
also revealed their distinct robustness characteristics.
In particular, we found that bright-bright solitons feature a robust propagation over long times.
On the other hand, as concerns solitons of the mixed-type (namely dark-bright and bright-dark ones),
we found that, in specific frequency bands (bands I-III), dark-bright solitons are more robust than those
in other bands (i.e., II-III) or bright-dark solitons: dark-bright solitons in bands II-III and bright dark solitons
preserve
%
%
their shape only for finite times and,
for sufficiently long evolutions, they are either
destroyed (bright-dark) or are significantly deformed (dark-bright).

We can thus postulate that from all types of solitons predicted analytically,
bright-bright and dark-bright ones (in bands I-III) are the most likely ones to be experimentally observable.
In all cases, our numerical results were found to corroborate the analytical
predictions, at least up to the times during which
the solitary waves propagate robustly.

It would be interesting to study other types of nonlinear CRLH-TL lattice models modelling
realistic structures composed by LH-metamaterials. In that regard, a pertinent interesting
direction would be the investigation of the effects of damping and driving, which may lead
to robust nonlinear waveforms which would constitute dynamical attractors
in such settings. Additionally, the study of higher-dimensional
settings is a particularly challenging problem. In the latter context,
in addition to simpler (yet genuinely higher dimensional, or even
quasi-one-dimensional) solitary wave structures, more complex waveforms
may be realizable such as vortices. The exploration of such states and
their dynamical robustness will be reported in future publications.

{\bf Acknowledgments.} The work of D.J.F. was partially supported by the Special Account of Research
Grants of the University of Athens. J.C. acknowledges financial support from the
MICINN project FIS2008-04848. PGK acknowledges support
from the US-NSF via CMMI-1000337, and the US-AFOSR via FA9550-12-1-0332.

\appendix
\section{The perturbation scheme}

Our analytical approximation relies on the use of the quasi-continuum approximation, which is a
variant of the method of multiple scales \cite{Jeffrey}. We introduce
new independent temporal variables, $t_n = \epsilon^n t$ ($n=0,1,2,\cdots$),
and accordingly expand the time derivative operator
$\partial_t$ as $\partial_t = \partial_{t_0} + \epsilon \partial_{t_1} + \ldots$.
Next, we seek solutions of Eq.~(\ref{eq:model}) in the form:
\begin{equation}
V_n =\sum_{\ell =1}\epsilon^{\ell} u_{\ell n}(t_n) +\cdots +{\rm c.c.}
\label{eq:ansatz01}
\end{equation}
Then, we substitute Eq.~(\ref{eq:ansatz01}) into Eq.~(\ref{eq:model}) and employ a continuum approximation for the envelope 
functions $u_n$, i.e., $u_n \rightarrow u(x)$, where $x=n\alpha$ and
$\alpha$ being the lattice spacing (the latter parameter does not appear in the results below, as one
may readily rescale $x$ as $x/\alpha$).Furthermore, we introduce the new spatial variables
$x_n = \epsilon^n x$ and, thus, $\partial_x = \partial_{x_0} + \epsilon \partial_{x_1} + \ldots$.
To this end, equating coefficients of like powers of $\epsilon$, we obtain the following (first three) perturbation equations:
\begin{eqnarray}
&\mathcal{O}(\epsilon ):& \hat{L}_0 u_1=0,
\label{eq:termS1}\\
&\mathcal{O}(\epsilon ^{2}):& \hat{L}_0 u_2+ \hat{L}_1u_1+ \hat{N}_0 u_1^2=0,
\label{eq:termS2}\\
&\mathcal{O}(\epsilon ^{3}):& \hat{L}_1 u_2+ \hat{L}_2u_1+\hat{N}_0[u_1u_2+\mu u_1^3]=0,
\label{eq:termS3}
\end{eqnarray}
where the operators are given by
\begin{eqnarray}
\hat{L}_0 &=&\frac{\partial^4}{\partial t_0^4}
+\left(1+\delta^2+4\beta^2\sin^{2}\frac{k}{2}\right)\frac{\partial^2}{\partial t_0^2}+\delta^2,\\
\label{eq:LO}
\hat{L}_1&=&4\frac{\partial^4}{\partial t_0^3 \partial t_1}
+2\left(1+\delta^2+4\beta^2\sin^{2}\frac{k}{2}\right)\frac{\partial^2}{\partial t_0 \partial t_1}\nonumber\\
&-&2i\beta^{2}\sin{k}\frac{\partial^3}{\partial t_0^3 \partial x_1}, \\
\label{eq:L1}
\hat{L}_2&=&\left(1+\delta^2+4\beta^2\sin^{2}\frac{k}{2}\right)\left(\frac{\partial^2}{\partial t_1^2}
+2\frac{\partial^2}{\partial t_0\partial t_2}\right)\nonumber\\
&-&6\frac{\partial^4}{\partial t_0^2 \partial t_1^2}
+4\frac{\partial^4}{\partial t_0^3 \partial t_2}\beta^2\cos{k}\frac{\partial^4}{\partial t_0^2 \partial x_1^2}\nonumber\\
&-&4i\beta^{2}\sin{k}\frac{\partial^3}{\partial t_0 \partial t_1 \partial x_1}-2i\beta^2\sin{k}\frac{\partial^4}{\partial t_0^2 \partial x_2^2},
\\ \label{eq:L2}
\hat{N}_0&=&\left(\frac{\partial^4}{\partial t_0^4}+\delta^2 \frac{\partial^2}{\partial t_0^2}\right).
\label{eq:N0}
\end{eqnarray}

We now seek for a solution of the linear problem, Eq.~(\ref{eq:termS1}), in the form:
\begin{equation}
u_1=\sum_{j=1}^{2}V_j(x_1,x_2,\ldots,t_1,t_2,\ldots)\exp(i\theta_j) +{\rm c.c.},
\label{eq:ansatzA}
\end{equation}
where subscripts $j=1$ and $j=2$ correspond to the LH and RH frequency bands,
$V_j$ is an unknown complex function, $\theta_j = k_j x_0-\omega_j t_0 $, while
the wavenumbers $k_j$ and frequencies $\omega_j$ satisfy the dispersion relation
provided in Eq.~(\ref{eq:linear_disp}).

Next, substituting Eq.~(\ref{eq:ansatzA}) into Eq.~(\ref{eq:termS2}), we obtain
the non-secularity condition for $l=1$:
\begin{equation}
\frac{\partial V_j}{\partial t_1}+\left[ \frac{\omega_j\beta^2\sin k_j}{2\omega_j^{2}-(1+\delta^{2}+4\beta^2\sin^{2}\frac{k_j}{2})}\right]\frac{\partial V_j}{\partial x_1}=0,
\label{eq:secular}
\end{equation}
which suggests that $V_j=V_j(X,x_2,\cdots, t_2,\ldots)$, where $X=x_1-v_{g_{j}} t_1$,
while the group velocities$v_{g_{j}}$ result self-consistently as
$v_{g_j}=\partial \omega_j/\partial k_j$ [cf. Eq.~(\ref{eq:gvelqc})].
Employing Eq.~(\ref{eq:secular}), we may determine from Eq.~(\ref{eq:termS2}), for $l=2$,
the unknown field $u_2$:
\begin{eqnarray}
&&u_2= -\sum_{j=1}^{2}\frac{4\omega_j^{2}(4\omega_j^{2}-\delta^{2})}{{{G_{j}}(2\omega_j,2k_j)}} V_j^2\exp(i2\theta_j)\nonumber\\
&&-\frac{2[(\omega_1+\omega_2)^{4}-\delta^2(\omega_1+\omega_2)^{2}]}{{{G_3}(\omega_1+\omega_2,k_1+k_2)}} V_1V_2\exp(i(\theta_1+\theta_2) \nonumber\\
&&-\frac{2[(\omega_1-\omega_2)^{4}-\delta^2(\omega_1-\omega_2)^{2}]}{{{G_4}(\omega_1-\omega_2,k_1-k_2)}} V_1V_2^{*}\exp(i(\theta_1-\theta_2)\nonumber\\
&&-\sum_{j=1}^{2}F_j(x_1,x_2,\cdots, t_1,t_2,\cdots)+{\rm c.c.},
\label{eq:u2}
\end{eqnarray}
where functions $G_j(\omega_j,k_j)$ are given by:
\begin{eqnarray}
G_j=&-&(1+\delta^2+4\beta^2\sin^{2}k_j)(2\omega_j)^{2}\nonumber\\
&+&(2\omega_j)^{4}+\delta^2,
\label{eq:coeff_G12}\\
G_3=&-&(1+\delta^2+4\beta^2\sin^{2}(\frac{k_1+k_2}{2}))(\omega_1+\omega_2)^{2}\nonumber\\
&+&(\omega_1+\omega_2)^{4}+\delta^2,
\label{eq:coeff_G3}\\
G_4=&-&(1+\delta^2+4\beta^2\sin^{2}(\frac{k_1-k_2}{2}))(\omega_1-\omega_2)^{2}\nonumber\\
&+&(\omega_1-\omega_2)^{4}+\delta^2.
\label{eq:coeff_G4}
\end{eqnarray}
On the other hand, functions $F_j(x_1,x_2,\cdots,t_1, t_2,\ldots)$ can be derived at the order
$\mathcal{O}(\epsilon ^{4})$, by means of the equation:
\begin{equation}
\hat{L}_2u_2+ \hat{N}_2 u_1^2=0,
\label{eq:term4}
\end{equation}
which leads to the result:
\begin{equation}
F_j=-\frac{2\omega_j^2\delta^2}{\omega_j^4+\delta^2}.
\label{eq:termF}
\end{equation}
To this end, we arrive at the following expression for $u_2$:
\begin{eqnarray}
&&u_2= -\sum_{j=1}^{2}c_j V_j^2\exp(i2\theta_j)-c_3 V_1V_2\exp(i(\theta_1+\theta_2) \nonumber\\
&&-c_4 V_1V_2^{*}\exp(i(\theta_1-\theta_2)-\sum_{j=1}^{2}c_{0j}|V_j|^2
+{\rm c.c.},
\label{eq:u22}
\end{eqnarray}
where
\begin{eqnarray}
c_j&=& \frac{4\omega_j^{2}(4\omega_j^{2}-\delta^{2})}{{{G_{j}}(2\omega_j,2k_j)}},\\
c_3&=&\frac{2[(\omega_1+\omega_2)^{4}-\delta^2(\omega_1+\omega_2)^{2}]}{{{G_3}(\omega_1+\omega_2,k_1+k_2)}},\\
c_4&=&\frac{2[(\omega_1-\omega_2)^{4}-\delta^2(\omega_1-\omega_2)^{2}]}{{{G_4}(\omega_1-\omega_2,k_1-k_2)}},\\
c_{0j}&=&\frac{2\omega_j^2\delta^2}{\omega_j^4+\delta^2}.
\label{eq:u2par}
\end{eqnarray}
Finally, defining the coefficients:
\begin{eqnarray}
A_j &=& c_{0j}+c_j,\\
B_{3-j}&=&c_{03-j}+c_3+c_4,
\label{eq:coeff_AB}
\end{eqnarray}
and using the variables $X=x_1-v_g t_1 \equiv \epsilon(n-v_g t)$ and $T=t_2 \equiv \epsilon^2 t$,
we derive from the non-secularity condition at $\mathcal{O}(\epsilon^{3})$ the coupled NLS equations (\ref{eq:MNLS1}).

\end{document}